\documentclass{aastex}
\usepackage{spr-astr-addons}
\usepackage{graphicx}   
\usepackage{multirow}
\usepackage{txfonts}
\usepackage{natbib}
\usepackage{url}

\begin{document}

\title{Accretion Flow Properties of GRS 1716-249 during its $2016-17$ `failed' Outburst}

\shorttitle{Flow Properties of GRS 1716-249}
\shortauthors{K. Chatterjee et al.}

\author{Kaushik Chatterjee\altaffilmark{1}, Dipak Debnath\altaffilmark{1$^*$}}
\affil{Indian Centre For Space Physics, 43 Chalantika, Garia Station Road, Kolkata, 700084, India \\ email (Dipak Debnath): dipakcsp@gmail.com}
\email{dipakcsp@gmail.com}
\and
\author{Debjit Chatterjee\altaffilmark{2,1}}
\affil{Indian Institute of Astrophysics, Koramangala, Bengaluru, Karnataka 560034}
\and
\author{Arghajit Jana\altaffilmark{3,1}} 
\affil{Physical Research Laboratory, Navrangpura, Ahmedabad 380009, India}
\and
\author{Sujoy Kumar Nath\altaffilmark{1}, Riya Bhowmick\altaffilmark{1}, Sandip K. Chakrabarti\altaffilmark{1}}
\affil{Indian Centre For Space Physics, 43 Chalantika, Garia Station Road, Kolkata, 700084, India}

\affil{$^1$Indian Centre For Space Physics, 43 Chalantika, Garia Station Road, Kolkata, 700084, India}
\affil{$^2$Indian Institute of Astrophysics, Koramangala, Bengaluru, Karnataka 560034}
\affil{$^3$Physical Research Laboratory, Navrangpura, Ahmedabad 380009, India}

\begin{abstract}

In $2016-17$, the Galactic transient black hole candidate GRS 1716-249 exhibited an outburst event after a long period of 
quiescence of almost 23 years. The source remained in the outbursting phase for $\sim 9$ months. We study the spectral 
and temporal properties of the source during this outburst using archival data from four astronomy satellites, namely MAXI, 
Swift, NuSTAR and AstroSat. Initial spectral analysis is done using combined disk black body and power-law models. 
For a better understanding of the accretion flow properties, we studied spectra with the physical two component advective 
flow (TCAF) model. Accretion flow parameters are extracted directly from the spectral fits with the TCAF model.
Low frequency quasi-periodic oscillations are also observed in the Swift/XRT and AstroSat/LAXPC data. From the nature 
of the variation of the spectral and temporal properties, we find the source remains in hard state during the entire outburst.
It never had a transition to other states which makes this event a `failed' outburst.  
An absence of the softer spectral states is consistent with the class of short orbital period objects, where the 
source belongs to. From the spectral fit, we also estimate the probable mass of GRS~1716-249 to be in 
the range of $4.50-5.93 M_\odot$ or $5.02^{+0.91}_{-0.52} M_\odot$.

\end{abstract}

\begin{keywords}
X-rays: binaries -- stars: black holes -- stars: individual (GRS 1716-249) -- accretion, accretion disc -- shock -- radiation
\end{keywords}

\section{Introduction}

Stellar-mass black holes (SBHs) are one of the most fascinating astronomical objects to study especially in X-ray, $\gamma$-ray. 
Black holes (BHs) are considered to be identified when the accretion flows around them emit radiations with spectra consistent 
with the black hole boundary conditions. Accretion disk are formed around the BHs when they accrete matter from the binary 
companion(s) through Roche-lobe overflow or wind. High energy radiation bears more information about the compact objects as 
they are emitted from closer radius of the accretion disk. Some of the stellar mass BHs are transient in nature, and show 
outbursts in a gap of months to years. An outburst is generally considered to be triggered, when rise in viscosity at the 
outer-edge of the disk occurs (Chakrabarti 1996; Ebisawa et al. 1996). According to Chakrabarti and his collaborators 
(Chakrabarti et al. 2019; Bhowmick et al. 2020), accreted matter from the companion first piles up at the far location 
(within Lagrange point L1), known as the pile-up radius ($X_p$) inside the Roche lobe. Due to the accumulation of large amount 
of matter at $X_p$, viscosity rises and crosses its critical value to trigger an outburst. During an outburst, a BH generally 
passes through four different spectral states. They are hard state (HS), hard intermediate state (HIMS), soft intermediate 
state (SIMS) and soft state (SS) (see, Remillard \& McClintock 2006; Nandi et al 2012). Normally, when an outburst starts, 
it begins in the HS. Then as the luminosity increases, it goes to the HIMS and then to SIMS. Finally it moves to the SS when 
luminosity of the source is the maximum. This is known as the rising phase of the outburst. Then, the luminosity starts to 
decrease slowly to the minimum and the source transits back to the HS in the reverse cycle. This is known as the declining 
phase of the outburst. In short, the spectral state transition of a BH takes place in the following sequence to form a hysteresis 
loop: HS (rising) $\rightarrow$ HIMS (rising) $\rightarrow$ SIMS (rising) $\rightarrow$ SS $\rightarrow$ SIMS (declining) 
$\rightarrow$ HIMS (declining) $\rightarrow$ HS (declining). 

Depending upon the nature of outbursts, transient black hole candidates (BHCs) show two types of outburst, such as type-I 
and type-II (Debnath et al. 2017). Those outbursts, where all four canonical spectral states are observed in the above 
mentioned sequence, are known as type-I or classical outburst. If any outburst does not go to softer states (SIMS, SS), 
it is termed as type-II or harder type of outburst. The later outbursts are also known as `failed' outbursts. Each of these 
states are defined based on the spectral and temporal properties. Hardness intensity diagram (HID; Belloni et al. 2005) 
and accretion rate ratio intensity diagram (ARRID; Jana et al. 2016) correlate the spectral and temporal properties during 
various spectral states. The radiation spectrum of a BH is defined by a multicolor thermal black body or disk black body 
(DBB) and a non-thermal power-law (PL) component. While harder states (HS, HIMS) are dominated by the non-thermal high energy 
radiations, the softer (SIMS, SS) states are dominated by thermal black body radiation. Quasi periodic oscillations (QPOs) 
are one of the most common features of a stellar mass black hole (SBH). Low frequency type C QPOs (van der Klis 2005) are 
most common which are generally observed in HS and HIMS (McClintock \& Remillard 2005). There are also the existence of 
sporadic type (A and B; Casella et al. 2005; van der Klis 2005) low frequency QPOs (LFQPOs). They generally show monotonic 
increase and decrease in frequency in rising and declining phases respectively. In SS, generally no QPOs are seen 
(Nandi et al. 2012; Debnath et al. 2010, 2013). 

Over the years, various models were introduced to understand physics of accretion around the BHs. Bondi flow (Bondi 1952), 
Standard disk model (Shakura \& Sunyaev 1973), Thick disk model (Paczynski \& Witta 1980) are a few of them. These models 
could explain the radiation spectra of a BHC successfully to some extent. However, they all were more of a special solution 
of a general picture. The soft component in the BH spectra is, as mentioned before, the black body radiation that comes from 
the Keplerian disk (Shakura \& Sunyaev 1973) and the PL component comes from the hot Compton cloud 
(Sunyaev \& Titarchuk 1980). In 1995, Chakrabarti and his collaborators came up with a general picture of accretion disk, 
namely the two component advective flow (TCAF) solution (Chakrabarti \& Titarchuk 1995, hereafter CT95; Chakrabarti 1997), 
which is a solution of radiation transfer equation considering both heating and cooling effects. In this TCAF model, there 
are two types of accretion flows: one is the high viscous, high angular momentum Keplerian matter, another one is the low 
viscous, lower angular momentum sub-Keplerian component. According to this model, the sub-Keplerian component, being less 
viscous, moves faster than the Keplerian one and forms an axisymmetric shock close to black hole when the centrifugal 
force roughly becomes comparable to the gravitational force acting on it. In the post-shock location as matter slows down 
and moves in a sub-sonic speed, matter starts to pile up and becomes hot. This post-shock puffed-up matter acts a `hot' 
Compton cloud as of Sunyaev \& Titarchuk (1980), known as the CENtrifugal pressure supported BOundary Layer (CENBOL). 
The soft photons from the Keplerian disk contribute to the soft multicolor thermal DBB part of the spectra. 
A fraction of these soft photons are intercepted by the CENBOL and are inverse-Comptonized multiple times by the `hot' 
electrons to become higher energy photons before they emit from the CENBOL. These Comptonized photons 
contribute to the non-thermal hard PL tail of the spectra. 

In general, at the beginning of an outburst, we see high dominance of the PL component. When an outburst is triggered 
due to a sudden rise in viscosity at the pile up radius i.e., at the outer disk, both types of matter start to move 
towards the BH after forming accretion disk (Ebisawa et al. 1996; Chakrabarti et al. 2019). 
As discussed before, being low viscous and low angular momentum, the sub-Keplerian matter moves closer to the BH much 
faster (roughly in a free fall time) than Keplerian component (moves in viscous time), and dominates at the beginning 
of an outburst. We see presence of a larger CENBOL. The source stays in HS during this phase of the outburst as sub-Keplerian 
matter dominates. The Keplerian component is referred to as the disk rate (${\dot m}_d$) while the sub-Keplerian component 
is referred to as the halo rate (${\dot m}_h$) in the TCAF model. As outburst progresses, more and more Keplerian matter 
comes into the picture to increase cooling rate, resulting shrinking of the CENBOL size.  
In this phase, we see the presence of the intermediate (HIMS and SIMS) spectral states. We observe 
soft spectral state, when CENBOL cools down completely due to higher dominance of the Keplerian matter. The declining 
phase starts, when viscous effect at the outer edge is turned off. This reduces supply of matter from the pile-up radius. 
In this phase, we see spectral states in the reverse sequence (SS$\rightarrow$SIMS$\rightarrow$HIMS$\rightarrow$HS) as 
supply from the ${\dot m}_d$ reduces in a faster rate compared to the ${\dot m}_h$. Debnath and his collaborators have 
successfully implemented this TCAF solution as a local additive table model in HEASARC's spectral analysis software 
package xspec to fit BH spectra (Debnath et al. 2014; 2015a). The model requires supply of initial guess of the four 
basic flow parameters: (i) Keplerian disk rate (${\dot m}_d$ in ${\dot m}_{Edd}$), (ii) sub-Keplerian halo rate 
(${\dot m}_h$ in ${\dot m}_{Edd}$), (iii) shock location ($X_s$ in Schwarzschild radius $r_s$), which is the boundary 
of the CENBOL, (iv) compression ratio ($R$), which is the ratio of the matter densities of the post to pre shock 
flows ($R={\rho}_+/{\rho}_-$), other than (v) mass of the BH ($M_{BH}$ in $M_\odot$) and (vi) normalization ($N$). 

In the HS, we generally see presence of a strong shock (higher $R$) at a larger $X_s$. Since the QPO frequency is 
inversely proportional to the location of the shock ($\nu_{qpo} \propto X_s^{-3/2}$; Chakrabarti \& Manickam 2000), 
we observe lower frequency QPOs in the HS. In the rising phase, as the outburst progresses, shock moves inward with 
reducing strength as post-shock cooling rate increases with increasing ${\dot m}_d$. We see monotonically increasing 
frequency of the generally observed type-C QPOs. On the HIMS to SIMS transition day, shock becomes weaker ($R \sim 1$) 
and we see highest frequency of the evolving type-C QPOs. These evolving QPOs are generally formed due to the resonance 
shock oscillation when cooling time scale roughly matches with the infall time scale (Molteni, Sponholtz \& Chakrabarti 1996;
Chakrabarti et al. 2015). In the SIMS, generally type-B or A QPOs are observed sporadically on and off. The origin of the SIMS 
QPOs are (i) due to the weak resonance of a weakly resonating CENBOL (for type-B) or a shockless centrifugal barrier (for 
type-A); or (ii) due to non-satisfaction of the Rankine-Hugoniot (R-H) conditions to form a stable shock (all types). Recently, 
Chatterjee et al. (2020) observed presence of type-C QPOs in the SIMS of MAXI~J1535-571 (using AstroSat LAXPC data), originating 
due to non-satisfaction of the R-H conditions. No QPOs are observed in SS. In declining intermediate and hard spectral states, 
we also see presence of the low frequency QPOs. Similar to the rising HS and HIMS, in the declining HS and HIMS, monotonically 
evolving type-C QPOS (with reducing frequencies as shock moves outward) are also generally seen. Jets are associated with the 
hard and intermediate spectral states. CENBOL acts as the base of the jets. Jets are more active in the intermediate spectral states. 
Compact jets are generally observed in the harder spectral states (HS and HIMS), while discrete or blobby jets could be seen in 
the SIMS. Due to the quenching of the CENBOL in the SS, similar to no QPOs, no jet is observed (Chakrabarti \& Nandi 2000 and references 
therein).

Analysis with TCAF model provides us a great understanding about the source's intrinsic property (i.e., mass) along with 
its spectral and temporal properties. Spectral fit with the TCAF model, directly provide us information about the flow 
dynamics via two accretion rates and two shock parameters. If mass of the BH is not well known, one can also estimate its 
probable value from spectral fit by keeping it as a free parameter. The normalization depends on the source's distance and 
inclination, which does not change significantly during an outburst. That is why $N$ generally stays almost constant, or, 
varies in a narrow range during an outburst. Because of the non-inclusion of the jet phenomena in the current version of the 
TCAF model \textit{fits} file, $N$ varies in a broad range in presence of jets. It can vary in a broad range if other 
phenomena like disk precision, bulk motion Comptonization, etc. are present. Recently, after its (TCAF) inclusion as 
an additive table model in XSPEC (Debnath et al. 2014, 2015a), spectra of many BHC have been successfully fitted and 
described by our group (Debnath et al. 2015a,b,2017,2020; Mondal et al. 2014, 2016; Chatterjee et al. 2016,2019; 
Jana et al. 2016,2020a,b; Chatterjee et al. 2020, Shang et al. 2019). Estimation of mass of these sources has also 
been done using the TCAF model. In our group's recent papers, X-ray contribution only from jets/outflows are also estimated 
using the method as described by Jana, Chakrabarti \& Debnath (2017, hereafter JCD17). 

The well known Galactic transient BHC GRS 1716-249 (also known as GRO J1719-24; Bharali et al. 2019) is an X-ray transient, 
which was discovered in 1993 by CGRO/BATSE and Granat/SIGMA telescopes (Ballet et al. 1993; Harmon et al. 1993). This 
source is located at a distance of $2.4 \pm 0.4$ kpc (della Valle, Mirabel \& Rodriguez 1994). Masetti et al. (1996) estimated 
the lower limit of the mass of the source, which is $\sim 4.9 M_\odot$. According to them, it has a K-type companion star 
of mass $\sim 1.6 M_\odot$ with an orbital period of $\sim 14.7$ hr. This BHC has a hydrogen column density of 
$\sim 0.4 \times 10^{22}~ cm^{-2}$ (Tanaka  1993). Since its last outburst in 1993, it was in a quiescence phase. In 2016, 
it became active and showed an outburst. On 2016 December 18, it was detected by MAXI (Masumitsu et al. 2016; Negoro et 
al. 2016). According to Negoro et al. (2016), during December 2016, the source had a photon index ($\Gamma$) 
$\sim 1.62 \pm 0.06$, which suggests that the source was in harder states during that time. Using Chandra satellite data, 
Miller et al. (2017) reported that during February 2017, $\Gamma$ still had a low value ($\sim 1.53$), which suggests 
that the source was still in harder state. Although Armas Padilla \& Munoz-Darias (2017) reported that the source was 
making a transition towards softer states during 2017 March 27 \& April 4, it was reported during the 2017 May 5 \& 11 
epoch that the source again returned back to the harder state (Bassi et al. 2017). According to them, the source did 
not go through the high soft state and hence they termed the $2016-17$ outburst of GRS 1716-249 as `failed' outburst. 

In this paper, we intend to study the accretion flow properties of the BHC GRS~1716-249 during its recent $2016-17$ 
outburst by careful analysis of the spectral and temporal properties and their evolution using the phenomenological 
disk black body plus power-law (DBB+PL) and physical TCAF models. The paper is organized in the following way. In \S2, 
we discuss about the data reduction and data analysis method. In \S3, we show the results from the spectral and 
temporal analysis. Finally, in \S4, we make discussions about the results and try to draw some conclusions from 
our work in \S5.

\section{Data Reduction and Analysis}

Swift monitored the $2016-17$ outburst of the BHC GRS 1716-279 roughly on a daily basis, starting from 2016 January 28, 
($\sim$ MJD 57781.00) till the end of the outburst. However, the outburst actually started in December. On 2016 December 18, 
MAXI detected this outburst (Masumitsu et al. 2016; Negoro et al. 2016). For our work, we make use of the archival 
Swift/XRT ($1-10$, $0.6-8$, $0.6-4$ and $4-10$ keV), MAXI/GSC ($2-10$ and $6-20$ keV) and Swift/BAT ($15-50$ and $15-30$ keV) data. 
Since we did not find the archival (both in the HEASARC and ASDC) Swift data before 2017 January 28, we make use of the 
on demand MAXI/GSC data from 2016 January 28. For broadband analysis, we also make use of the AstroSat/LAXPC data from the 
ISSDC archive and NuSTAR data from HEASARC archive. The data reduction for different satellites/instruments are done 
using the methods mentioned in the following sub-section.

\subsection{Data Reduction}

\vspace*{0.3cm}
\textbf{Swift/XRT} 

We make use of the windowed timing (WT) mode data for XRT instrument. We use the \textit{xrtpipeline} command to produce 
cleaned event files from the level-1 uncleaned data. For each cleaned event file, we select and save a circular region of 
30 arcsec around the source in $ds9$ using the XSELECT task. We also select and save a background region, away from the 
source. Using these region files, we extract spectra for both the source and the background by using the methods
described in XRT data analysis thread (\url{https://www.swift.ac.uk/analysis/xrt}). After producing the spectra, we make 
use of the \textit{grppha} task with a minimum of 10 counts/bin. The default time bin in the XRT event files is $1~sec$. 
For our analysis to find LFQPOs, we extract light curves of bin size $0.01~sec$ and for that, we use the 
`\textit{set binsize 0.01}' command in XSELECT. 

\vspace*{0.3cm}
\textbf{Swift/BAT} 

For the reduction of BAT archival data, we follow the methods mentioned in BAT analysis thread 
(\url{https://www.swift.ac.uk/analysis/bat}). At first, we use \textit{bateconvert} to make proper energy conversion 
from the already present event file. Next, \textit{batbinevt} is used to produce proper detector plane image (DPI). 
After that, we use \textit{bathotpix} to define hot pixels and \textit{batmaskwtevt} to apply mask weighting. 
With inclusion of the mask weighting file to the \textit{batbinevt} command again, we produce the spectra files. 
Then, we use \textit{batphasyserr} to add systematic error to the spectra files. After that, ray tracing correction 
is done by using the \textit{batupdatephakw} command. At last, we use \textit{batdrmgen} to generate the response file. 
We have not made any light curves from the BAT data. 

\vspace*{0.3cm}
\textbf{MAXI/GSC} 

The MAXI/GSC spectra files are available on demand in the website (\url{http://maxi.riken.jp/mxondem}) and we 
downloaded suitable spectra files which are close to or on the XRT data dates. The on-demand process is described 
in Matsuoka et al. (2009).

\vspace*{0.3cm}
\textbf{AstroSat/LAXPC} 

For extracting LAXPC data, we use publicly available code (http://www.tifr.res.in/$\sim$antia/laxpc.html). 
We perform the data reduction process to generate level-2 files to create source and background spectra for LAXPC10 unit 
(unit 1) using all anode layers. First, we run the `\textit{laxpcl1.f}' program to process multiple orbits of the level-1 
data and produce event files, light curves, spectra, good time intervals (GTI) in both the ASCII and FITS format. Then we 
run the `\textit{backshiftv3.f}' program to make background correction to the light curves and spectra and also to identify 
the response file, to be used in the spectral analysis. We also extract $0.01$~sec light curves for QPO analysis.

\vspace*{0.3cm}
\textbf{NuSTAR/FPMA} 

For getting broad energy information ($3-79$~keV), we use FPMA instrument data of the NuSTAR satellite. The data 
reduction process is done using the NuSTAR data analysis software \textit{NuSTARDAS} (version 1.4.1). We first run 
the NuSTAR pipeline using \textit{nupipeline} command using the uncleaned stage-I data providing input, output 
directories and steminputs. It produces stage-II cleaned event files. Using XSELECT, we read the cleaned event files 
and save the source and background region files. Now, using those region files, we use the \textit{nuproducts} command 
to extract stage-III data to produce background subtracted spectra. We also made use of the \textit{grppha} task for 
grouping the data with a minimum of $30$~counts/bin.

\subsection{Data Analysis}

\subsubsection{Temporal Analysis}

The daily average MAXI/GSC ($2-10$~keV) and Swift/BAT ($15-50$~keV) light curves are downloaded from the respective 
public archives. Those light curves are converted in \textit{mCrab} unit using proper conversion factors. For GSC, 
1~mCrab is equal to 0.00282 $photons~cm^{-2}~s^{-1}$ in $2-10$ keV, while for BAT, 1~mCrab equals 0.000220 
$counts~cm^{-2}~s^{-1}$ in $15-50$ keV. We also estimate the count rates in $0.6-4$~keV and $4-10$~keV energy bands 
from XRT light curves.

For QPO analysis, we generate the power density spectra (PDS) by running the \textit{powspec} task of the XRONOS package. 
We extract QPO information from both the XRT and the LAXPC data using $0.01~sec$ time binned light curves. For XRT, we use 
$1-10$~keV light curves, while for LAXPC, we use $3-80$ keV light curves. Geometrical rebinning constants of `$-1.02$' or 
`$-1.05$' (as needed) has been used. To fit the QPO profiles, we make use of the Lorentzian model. This gives us the fitted 
value of QPO frequencies ($\nu_{qpo}$), its width ($\triangle \nu$) and power. With these, we calculate the quality factor 
($Q=\nu_{qpo}/\triangle\nu$), rms ($\%$) value, which help us to classify the nature (type) of the QPOs. 

We have also produced LAXPC light curves' data in four different energy bands, which are $3-20$ keV, $20-40$ keV, $40-60$ keV
and $60-80$ keV respectively. By generating the light curves in the mentioned energy bands, we have produced PDS for those 
ranges to see the energy dependence of QPOs.

\subsubsection{Spectral Analysis}

We have fitted spectra using XRT, GSC, BAT, LAXPC, and NuSTAR data in different energy ranges. Data of a total of 39 
observations are fitted using both the phenomenological disk blackbody plus power-law (DBB+PL) and physical TCAF model. 
10 observations are fitted using only XRT data in $0.6-8$~keV band, while 25 observations are fitted using XRT and GSC data 
in the range of $0.6-20$ keV and only 2 observations are fitted using XRT, GSC and BAT data in the energy range of 
$0.6-30$ keV. Since, there is a significant noise in the BAT data beyond $30$~keV, we have only taken them in the $15-30$ keV 
range. For broadband fitting, we have used LAXPC and NuSTAR and combined them with the XRT data (of same MJDs). Using 
combined XRT-LAXPC data, we have fitted 2 observations (out of the total 39 studied observations) in $0.6-80$ keV energy 
band on MJDs $\sim$ 57799 and 57850 respectively. On the other hand, we have fitted 2 (out of those 39) observations using 
combined XRT-NuSTAR in $0.6-77$ keV energy band on MJDs $\sim$ 57850 and 57853 respectively. We have used \textit{TBABS} 
(Wilms, Allen \& McCray 2000) model to account for the absorption in the interstellar medium. We use \textit{TCAF} as an 
additive table model along with the multiplicative \textit{TBABS} model. $1 \%$ systematic error is used for all our spectral 
analysis except for the AstroSat/LAXPC data, for which we use $2 \%$ systematic error (Antia et al. 2017; 
Sreehari et al. 2019). 

\section{Results}

We have studied the accretion flow properties of the source based on the temporal and spectral analysis. Because every spectrum 
is fundamentally dependent on the mass of the black hole which controls the thermodynamic quantities of a disk, it is
possible to estimate the mass of the black hole from each spectrum if the fit is proper. TCAF fits the data in absolute
sense up to a normalization (a function of distance). Thus we have also obtained the probable mass of the source 
from TCAF model while fitting of the spectra. Our results are presented below.

\subsection{Temporal Properties}

Here, we discuss about the timing properties (light curves, hardness-ratios, QPOs) and their evolution over the duration 
of the outburst. 

\subsubsection{Evolution of Outburst Profiles}

In Figure 1, we show the entire duration of the $2016-17$ outburst of GRS~1716-249. The outburst started after MJD 
$\sim 57710$ (2016 November 18) and was continued for almost 9 months. The source returned back to the quiescence close 
to MJD $\sim$ 58000 (2017 September 4). In panel (a) of Figure 1, we show the $2-10$ keV MAXI/GSC (online red) and 
$15-50$ keV Swift/BAT (online blue) fluxes from MJD $\sim$ 57720 (2016 November 28) to 58003 (2017 September 07). 
According to Debnath et al. (2010), depending upon the variation of flux or count rates in light curve profiles, 
outbursts are divided into two types : \textit{fast rise slow decay} (FRSD) and \textit{slow rise slow decay} (SRSD). 
From GSC and BAT light curve profiles in Fig. 1(a), we notice that both the fluxes increase significantly in a very 
short time of the initial rising phase of the outburst. For BAT flux (online blue), it was $\sim$ 59.9 mCrab on 
MJD $\sim$ 57720 (2016 November 28) and in a short period over 50 days, it reaches its peak of $\sim$ 604.5 mCrab 
on MJD $\sim$ 57772 (2017 January 19). Then it started decreasing slowly and went back to its quiescence after a period 
of almost 8 months. For GSC flux (online red), the situation is similar. It started increasing after MJD $\sim$ 57720 
(2016 November 28) on which it has a flux of $\sim$ 20.4 mCrab. It increased from this rapidly to a flux of 
$\sim$ 465.4 mCrab on MJD $\sim$ 57768 (2017 January 15) and then decreased to a value of 355.7 mCrab on 
MJD $\sim$ 57784 (2017 January 31). It started increasing and gradually reached its peak of 491.9 mCrab on 
MJD $\sim$ 57823 2017 March 11). Then it decreased and reached quiescence after MJD $\sim$ 58000 (2017 September 04). 
Depending on these variations, the nature of variation of the outburst can be classified as `FRSD' type. During the entire 
outburst, the hard (BAT) flux was dominant over the soft (GSC) flux. In Fig. 1(c), we show the XRT count rates 
in two bands: $0.6-4$ keV (online red), and $4-10$ keV (online blue). The $4-10$ keV count rate is multiplied by a factor 
of 4 to scale it with the $0.6-4$ keV count rate. Due to absence of Swift XRT data during the initial phase of the outburst, 
we are unable to show the full variation of XRT count rates. The count rates are estimated from those observations for which 
we have performed spectral analysis. 

\subsubsection{Hardness Ratio}

Hardness ratio (HR) is defined as $F_H/F_S$, where $F_H$ and $F_S$ represent the hard and soft X-ray fluxes (or the count rates) 
respectively. It provides us with a quick look on evolution of the flow dynamics of the source, as it is assumed that origin 
of the $F_H$ is non-thermal and $F_S$ is thermal processes. But, from the physical point of view, there is always a chance 
of mixing up of the thermal and non-thermal photons while $F_H$ and $F_S$ are defined as photon flux or count rates into two 
fixed energy bands. In the harder states (HS \& HIMS), the HR is generally high, when the contribution from the hard 
(non-thermal) component dominates over the soft (thermal) component. In softer states (SIMS \& SS) it becomes low when the 
soft component takes over hard component.

In Figure 1, we have shown two HRs using the ratio of $15-50$ keV BAT hard flux to $2-10$ keV GSC flux (HR1; panel b) and 
$4-10$ keV XRT to $0.6-4$ keV XRT count rates (HR2; panel d). Since, there is absence of XRT data in the HEASARC archive in 
the rising phase, we only show the part of the HR2 that we use for our spectral analysis, i.e., from MJD 57781 (2017 January 
28). From panel b, we notice that at the beginning of the outburst, HR1 was around $\sim$ 1.5 and slowly decreased to a value 
of $\sim$ 1 on MJD $\sim$ 57855 (2017 April 12). Then it stayed on this value for some time and then reached to a minimum 
of 0.5 on MJD 57959 (2017 July 25) and then again increased gradually afterwards. From this variation, we can suggest 
that the source did not go through all the defined spectral states of a BH. Clearly, any sign of softer states is missing 
from this HR. HR2 does not also suggest any strong evidence of the presence of softer states. From the point of view of 
hardness ratio, we could say that the source has gone through only the harder states (HS, HIMS). However, we need to do 
spectral analysis to draw a firm conclusion on the spectral states.

\subsubsection{QPO Evolution}

As described in \S2, we have made use of the 0.01 sec time binned Swift/XRT (in $1-10$ keV) and AstroSat/LAXPC (in $3-80$ keV) 
light curves to produce PDS using XRONOS package. We show a PDS (in $0.2-10$ Hz) in Fig. 2. The continuum of PDS (Fig. 2) is 
fitted using combination of Lorentzian and power-law profiles. Using Lorentzian model, we have fitted all the light curves' 
data which show QPO nature and extracted their frequencies. We have found a total of 6 days (from both the XRT and LAXPC data) 
on which QPO was present. In Fig. 3, we show the QPO frequencies of these 6 days along with their evolution with time during 
the outburst. We see monotonic increase of QPO frequency during the outburst's rising phase. On MJD 57781.00 (2017 January 28), 
we see the presence of the first QPO with $\nu_{qpo} =0.633$ Hz. Then, $\nu_{qpo}$ increases to 0.675 Hz on MJD 57799.68 (2017 
February 15). On this day, due to noise (i.e., low signal to noise ratio), we have not found any QPO nature in the XRT data. 
Then $\nu_{qpo}$ kept on increasing to reach to a value of 1.213 Hz. This is also from the LAXPC data. We have not found any 
sign of QPOs in the declining phase of the outburst. The evolution is shown in Figure 3. The detailed QPO information is provided 
in Table 1. 

According to Chakrabarti et al. (2015), in the TCAF model, when the cooling timescale (time, the soft disk photons take to cool 
the CENBOL in the process of inverse-Comptonization or by synchrotron cooling if magnetic field is strong) roughly ($\sim 50 \%$)
matches the infall timescale (time taken by matter to fall radially inside from the post-shock region), the resonance condition 
for the shock oscillation is satisfied. Here for this outburst, we have also estimated these timescales. During the entire phase 
of the outburst, these two timescales were not of the same order. Resonance for shock oscillation did not satisfy in both 
the rising and declining phases. 

The nature or type of QPOs are classified according to Q-value, rms ($\%$) amplitude (see, Casella et al. 2005). We show 
these estimated properties for these 6 QPOs in Table 1. We classified observed QPOs as type C and B.

\subsubsection{Energy Dependent PDS}

Using the AstroSat/LAXPC data of 2017 February 15 (MJD 57799.68) and 2017 April 06 (MJD 57849.69), we studied energy 
dependent nature of the observed QPOs. We have produced the power density spectra (PDS) in four energy bands: $3-20$, 
$20-40$, $40-60$ and $60-80$ keV to check the energy dependence of the QPOs. In Fig. 4, we show model fitted PDS in 
between $0.2-10$ Hz in the above four energy bands for the data observed on 2017 April 06 (MJD 57849.69; obs id = 
164T01-9000001140; orbit = 8238). From the PDS, there is a clear signature of the primary QPO ($\sim 1.2$~Hz) and its 
harmonics ($\sim 2.4$~Hz) in the two lower ($3-20$ keV and $20-40$ keV) energy bands (see, Fig. 4(a-b)), whereas in 
$40-60$ keV band, we see only the primary QPO. In the $60-80$ keV band, signatures of the primary QPO becomes weaker. 
But, in the the two higher energy bands ($>40$~keV), we see the signature of another low frequency QPO at $\sim 0.62$~Hz. 
This low frequency QPO is more prominent in the $60-80$ keV band. From the Lorentzian fitted QPOs in the PDS, we have 
estimated the rms of the primary QPO in the $3-20$, $20-40$ and $40-60$ keV energy bands. A decreasing trend of the 
rms of the primary QPO is observed as we moved toward higher energy bands, this means that QPO becomes fainter as we 
move towards the higher energy band. The QPO rms is found $\sim 12$ percent in $3-20$ keV, $\sim 11$ percent 
in $20-40$ keV, and $\sim 7$ percent in $40-60$ keV. These results are presented in Table 1 (bottom panel).

\subsection{Spectral Properties}

The study of the spectral properties of an outburst is an important aspect to get an idea about the source and its evolution 
during the outburst. First, we have fitted 39 observations with the \textit{DBB+PL} model, using mainly Swift/XRT, MAXI/GSC 
and Swift/BAT data. The \textit{DBB+PL} fitting gives a rough approximation about the spectral states. To understand the 
physical picture of accretion flow properties, we have refitted those same observations with the physical \textit{TCAF} 
model. This shows us the evolution of the accretion flow parameters and also gives us the estimation of the mass of the 
source. This will be discussed in a different subsection. To check consistency of our result,  we have fitted two 
observations in broader bands with each of NuSTAR/FPMA and AstroSat/LAXPC data by combining them with simultaneous Swift/XRT 
data. We have fitted XRT+NuSTAR and XRT+LAXPC data in $0.6-77$ keV and $0.6-80$ keV energy bands respectively. We show 
detailed spectral analysis results in the following section. In Fig. 5, we show four \textit{TCAF} model fitted spectra 
using different spectral data in different energy bands. In panel (a), we show the spectra using XRT+GSC+BAT instruments 
in the $0.6-30$ keV band, in (b) XRT+GSC spectra in $0.6-20$ keV is shown. In Fig. 5(c-d), we show the fitted spectra 
in the broader bands using simultaneous XRT+NuSTAR ($0.6-77$ keV) and XRT+LAXPC ($0.6-80$ keV) data respectively.

We show the variation of \textit{TCAF} and \textit{DBB+PL} model fitted parameters in Figs. 6 and 7. In Figure 6, we 
show the variations of TCAF model fitted (a) total accretion rate (${\dot m}_d + {\dot m}_h$), (b) disk rate (${\dot m}_d$), 
(c) halo rate (${\dot m}_h$) and (d) accretion rate ratio (ARR), which is defined as the ratio of halo rate to discrete 
(${\dot m}_h/{\dot m}_d$). Both of these rates are in the unit of Eddington rate (${\dot m}_{Edd}$). We show the shock 
parameters ($X_s$ and $R$) of \textit{TCAF} model in Fig. 7(a-b). The \textit{DBB+PL} model fitted inner-disk temperature 
($T_{in}$) and photon index of power-law ($\Gamma$) are shown in panel c-d of Fig. 7.

We notice that from the beginning of our analysis date up to MJD 57827.78 (2017 March 15), ${\dot m}_h$ remained at a nearly 
constant value of $\sim 0.48$. Then it decreased by a very small amount to reach to a value of $0.45$ on MJD 57878.74 
(2017 May 05) and then it gradually decreased and reached a value of 0.20 on MJD 57975.27 (2017 August 11). From the 
beginning, ${\dot m}_d$ was gradually increasing from a value of 0.23 and reached its maximum value of 0.37 on MJD 57857.02 
(2017 April 14). The total flow rate (${\dot m}_d + {\dot m}_h$) also showed similar variation with ${\dot m}_d$. 
The total rate was $0.72$ on the first day of our observation and after that it gradually increased, and reached its peak 
on MJD 57857.02. After that, both the disk rate and total rate started decreasing and reached their minimum 
($\sim 0.12 ~\&~ 0.32$) on MJD 57975.27. The shock location was far away (Fig. 7a) at a distance of $\sim 240 r_s$ 
on the first day of our observation on MJD 57781.00 (2017 January 28). After MJD 57786.73 (2017 February 02; on which 
it became $\sim 236 r_s$), and it started decreasing gradually and attained its minimum of $\sim 150 r_s$ on MJD 57796.55 
(2017 February 12). It remained at this minimum distance for some time till MJD 57857.02 (2017 April 14) before increasing 
gradually. On the last day (MJD 57975.27; 2017 August 11) of our analyzed period, the shock ($X_s$) moved away to a distance 
of $\sim 246 r_s$. The compression ratio (ratio of post to pre shock density; $\rho_+/\rho_-$; Fig. 7b) has varied between 
$2.3-3.4$. The inner-disk temperature ($T_{in}$) was $\sim 0.74$ keV on MJD 57781.00. It first increased to $\sim 0.84$ keV 
on MJD 57790.46 (2017 February 06) and then gradually decreased to $\sim 0.53$ keV on MJD 57857.02. Then it again increased 
to $\sim 0.70$ keV on MJD 57884.06 (2017 May 11) after which it decreased gradually. The power-law photon index ($\Gamma$)  
varied between $1.25-1.76$. On first observation $\Gamma$ was low ($\sim 1.3$). Then it gradually increased and attained 
its maximum value ($\sim 1.76$) on MJD 57857.02, the day on which the total rate was maximum. It then started decreasing 
and became $\sim 1.42$ on MJD 57975.27 (2017 August 11).

As we discussed before, from our temporal results one could conclude that the source might have gone through the HS and HIMS. 
However, our spectral results do not strongly support the presence of the HIMS. We see a complete dominance of the halo rate 
(${\dot m}_h$) over the disk rate (${\dot m}_d$) throughout the entire duration of our analysis. The photon index ($\Gamma$) 
never became $> 1.8$ in any phase of the outburst. However, both of our temporal and spectral results support the absence of 
softer states during this outburst. Based on our spectral and temporal results, we can surmise that the source GRS~1716-249 
went through only the hard state during its recent $2016-17$ outburst. 

We also show the variations of the hydrogen column density ($n_H$) (Fig. 8a), which is found to be between $0.37$ and $0.61  
\times 10^{22} ~cm^{-2}$ using the \textit{TBABS} model. This result is in good agreement with Tanaka (1993). 

As we mentioned earlier that the normalization parameter ($N$) of the TCAF model is a constant multiplicative factor, 
to overlay the observed spectra to the theoretical spectra of the TCAF model \textit{fits} (Molla et al. 2016, 2017). In our 
analysis, the variation of normalization ($N$) is shown in Fig. 8(b), which varied in a range of $12.99-29.70$. There is 
report on the presence of radio jet during this outburst by Bassi et al. (2019). So, this broad range of variation of $N$ 
is due to the presence of radio jet.

\vspace{0.2cm}
\subsection{Estimation of Mass from Spectral Analysis}

Any consistent theory of accretion flows would show that the thermodynamic parameters, such as the temperature,
density etc. of the flow is inherently a function of the black hole mass. Since the radiations emitted 
carry the information of the density and temperature distribution, the spectrum in the absolute terms
must contain the information of the mass of the black hole. Although the phenomenological `DBB+PL' model
fits the spectral shape, there is no information about the mass. A physical model such as TCAF is capable of 
extracting the mass, though the accuracy is restricted as the observed spectrum is a sum of radiation contributed 
from a large region of varied density and temperature.

To fit black hole spectrum, TCAF model needs to supply six input parameters including mass ($M_{BH}$) and normalization 
and the rest are related to the flow parameters. When we fit a spectrum the TCAF model parameters after keeping all 
parameters as free, we get an estimation of the best fitted values for all those parameters. We keep mass as a free 
parameter when it is unknown. While fitting the spectra of GRS~1716-249, we obtained it to lie between $4.50-5.93 M_\odot$ 
from the fits. The uncertainty arises from non-uniformity in the quality of the data and errors due to fitting. Our 
estimation of the probable mass of the source from the TCAF analysis is $5.02^{+0.91}_{-0.52} M_\odot$ (since $5.02~M_\odot$ 
is average of the model fitted mass values). This is in good agreement with the previous finding of Masetti et al. (1996), 
who estimated the mass of the source as $4.9 M_\odot$. In Figure 8(c), we show our model fitted mass values throughout 
the entire analysis period of the outburst.

\vspace{0.5cm}
\section{Discussions}

The Galactic transient BHC GRS 1716-249 was monitored in multi-wave band by Swift roughly on a daily basis starting from 
MJD 57781.00 (2017 January 28). The data of the initial rising phase of the 2016-17 outburst is missing as Swift started 
pointing the source few days after its discovery. MAXI also monitored the outburst. We have performed spectral analysis 
of GRS 1716-249 during its $2016-17$ outburst using archival data of Swift/XRT ($0.6-8$ keV), MAXI/GSC ($6-20$ keV) and 
Swift/BAT ($15-30$ keV). To study in the higher energy bands, we use two observations of NuSTAR (on MJD 57850.60; 
2017 April 07 and 57853.69; 2017 April 10) and also two observations of AstroSat/LAXPC (on MJD 57799.68; 2017 February 15 
and 57849.69; 2017 April 06). Combining NuSTAR or LAXPC data with simultaneous XRT observations, a broad energy study is 
done in the range of $0.6-77$ keV and $0.6-80$ keV bands respectively. The spectral analysis is done using both phenomenological 
(DBB+PL) and physical (TCAF) models. The timing analysis is done mainly using XRT ($1-10$~keV) and LAXPC ($3-80$~keV) data. 

Low frequency QPOs are one of the most common phenomena in hard and intermediate spectral states of stellar mass BHCs. 
For QPO analysis, we have made use of $0.01$~sec time binned light curves from both XRT and LAXPC data in $1-10$~keV 
and $3-80$~keV energy bands respectively. With Lorentzian model fitting, we have extracted the centroid frequencies 
of the QPOs. Using XRT data, we have found QPOs in 4 observations on MJDs 57781.00 (2017 January 28; obs. Id.=00034924001), 
57806.57 (2017 February 22; obs. Id.=00034924019), 57808.37 (2017 February 24; obs. Id.=00034924020) and 57821.53 
(2017 March 09; obs. Id.=00034924022). The power density spectra, made from XRT data, has low signal to noise ratio. 
We think that there might be existence of LFQPOs on other dates where no prominent QPOs are observed due to the noise. 
We also checked PDS from LAXPC data on MJDs 57799.68 (2017 February 15) and 57849.69 (2017 April 06). On both these 
observations, we found existence of QPOs. The PDS from LAXPC data are much less noisy than that of XRT. In Fig. 2, 
we show model fitted PDS of the first LAXPC observation in the frequency band $0.2-10$ Hz. This Figure shows a QPO 
with a harmonic, which are fitted using the combination of Lorentzian and power-law models. Using all the QPO information 
from both set of data, we show that $\nu_{qpo}$ increased monotonically in the rising phase (see, Fig. 3). We have not 
found the existence of QPOs in the declining phase. To find the origin of these QPOs i.e., they are due to satisfaction 
of the resonance oscillation of the shock or not (see, Molteni et al. 1996; Chakrabarti et al. 2015), we calculated cooling 
time ($t_c$)and infall time ($t_i$) scales of the post shock matter and found that their ratio ($t_c/t_i$) deviates largely 
from the unity. So, we conclude that during the both rising and declining phases of the outburst, resonance shock condition 
was not satisfied. Possibly due to this, in the declining phase of the outburst, we have not observed any QPO.

We have also shown the energy dependence of power density spectrum using AstroSat/LAXPC data. We have observed the presence 
of both the fundamental QPO and the harmonic appear in energies $<40$ keV, while there is no sign of harmonic above 40
keV. The QPO signature becomes weaker in the PDS above 60 keV. It is evident from the Figure 4(a-d) that as the energy 
is increasing the QPO shape is changing along with reducing rms. With increasing energy, the QPO rms is decreasing while 
the broadband noise is increasing. The disappearance of QPO shape in high energy (above 60 keV here) in the LAXPC data 
could be due to the reason that there is presence of low number of photons in high energy. It also could be due to the
fact that the detector effective area decreases to $4100-2200~cm^2$ (Antia et al. 2017) in the energy range of $60-80$ keV. 
This was also reported by Sreehari et al. (2019) for the BHC MAXI J1535-571. Interestingly, we see one more low frequency 
QPO $\sim 0.62$~Hz in the energy range above $40$~keV. This QPO becomes more significant in the $60-80$~keV band as compared 
to $40-60$~keV band. This has an rms of $\sim 3.6\%$ in $60-80$ keV band.This could be one more local oscillation in the higher 
radius. 
 
From the spectral analysis with the TCAF model, we see that at the beginning of the outburst, the supply of both the 
high viscous Keplerian and low viscous sub-Keplerian matter stays low. According to the TCAF model, due to higher radial 
velocity of the sub-Keplerian component, it moves much faster than the Keplerian component (moves in viscous timescale). 
The Keplerian component, having high angular momentum forms the disk, with sub-Keplerian matter residing over it, forming 
the halo. This halo component forms the hot Compton cloud or CENBOL at the post-shock region. Due to inverse Comptonization, 
the soft photons from the disk gains energy and produced the power-law tail in the observed spectrum. At the beginning of 
the outburst, the shock was located far away ($\sim 240 r_s$) and the cooling was inefficient. As a result the shock was 
strong. The shock then gradually moved closer and reached $\sim 150 r_s$ after MJD $\sim$ 57800 (2017 February 16). 
This means that the Keplerian matter moved closer and the cooling was increased. During this time, the total accretion 
(disk + halo ) rate was increased. Although the halo rate was decreasing from the start, it did not decrease that much 
during this period of observations (from $\sim 0.49$ to $\sim 0.47$). While the disk rate steadily increased 
 during this period, the halo rate was always dominating. Also, during this time, the Photon index ($\Gamma$) 
was also on the lower side ($1.25-1.66$) of its values, with $T_{in}$ decreasing from $0.74$~keV to $0.58$~keV gradually.
The ARR (Fig. 6d) was also higher ($\sim 1.4-2.1$) during this period of the outburst. Since the total rate continued 
to increase in this period, we term it as the rising phase. The shock location was stalling at $\sim 150 r_s$ after 
2017 April 07 (MJD 57850.37) for some days until 2017 April 21 (MJD 57864.07). From MJD=57864.07, we observe outward 
movement of the shock, when the total rate was maximum. The disk rate also reached to its peak just one day prior to this 
on 2017 April 14 (MJD 57857.02). It was still near its peak on MJD=57864.07. This is when we believe 
that the declining phase started. After this day, the disk rate started to decrease with already decreasing halo rate. 
Due to this, the efficiency of cooling of the CENBOL decreased and the boundary i.e., $X_s$ started to move outward. 
On 2017 May 05 (MJD 57878.74), $X_s$ became $\sim 194 r_s$ and then moved away rapidly from the source. On MJD=57857.02, 
the photon index ($\Gamma$) of power-law model was on its highest value ($\sim 1.76$). ARR became $\sim 1.3$ during this 
period. After MJD=57878.74, as the supply of Keplerian matter decreased further, the $X_s$ continued to move outside and 
reached to a similar value ($\sim 240 r_s$) as in the beginning of the outburst. After MJD=57878.74, $\Gamma$ 
again became $<1.6$ and both the ${\dot m}_d$ and ${\dot m}_h$ further decreased, as the source progressed towards the 
quiescence. ARR did not increase that much except only at the back end of the outburst when it became $\sim 1.7$ on 
MJD=57975.27 (2017 August 11). Since the supply of the Keplerian rate did not become very high during the outburst, 
the CENBOL was not cooled down completely during the middle phase of the outburst. As a result it started to move outside 
when supply resides and ultimately source did not reach intermediate and softer states. Due to the absence of softer states, 
the $2016-17$ outburst of GRS~1716-249 can be referred to as a harder type (type-II) or `failed'. According to 
Bharali et al. (2019), the orbital period of the source is $14.7$~hrs. So, this nature of the non-observation of the 
softer spectral states is consistent with other short orbital period objects, such as MAXI J1836-194 (Jana et al. 2016), 
Swift~J1753.5-0127 (Debnath et al. 2017), XTE~J1118+480 (Chatterjee et al. 2019). A similar presence of only HS during 
2000 \& 2005 outbursts of the short orbital period ($\sim 4.1$~hrs)  transient BHC XTE~J1118+480 was noticed 
(see, Chatterjee et al. 2019; Debnath et al. 2020). 

We also estimated the probable mass of the BH ($M_{BH}$), from our spectral analysis with the TCAF model. In TCAF, $M_{BH}$ 
is a model input parameter. Since, mass of GRS~1716-249 is not well known, while fitting spectra we kept $M_{BH}$ as free. 
We observed $M_{BH}$ in the range of the source in between $4.50-5.93 M_\odot$. The observed variation of our estimated 
mass is mainly due to instrumental and fitting errors. The flux from black body emission depends on the fourth power of the 
temperature ($T$) of the Keplerian disk. So any small change in the value of $T$ while fitting the black body component
(which is possible due to poorly understood absorption due to intervening medium) can lead to a significant change in the 
estimated mass. Taking the average, we conclude that the most probable mass of the BHC GRS 1716-249 is $5.02^{+0.91}_{-0.52} 
M_\odot$, which is in good agreement with the earlier findings (Masetti et al. 1996). 

\section{Summary and Conclusions}

We have done spectral and temporal analysis of the BHC GRS 1716-249 during its recent $2016-17$ outburst. To study outburst 
profile in different energy bands and hardness ratios, we used MAXI/GSC, Swift/BAT and Swift/XRT data. The temporal analysis 
of the QPOs is done using the Swift/XRT and AstroSat/LAXPC data using \textit{powspec} of \textit{XRONOS} package. We have 
also studied energy dependent QPOs using the LAXPC data. The spectral analysis is done using Swift/XRT, MAXI/GSC, Swift/BAT, 
AstroSat/LAXPC and NuSTAR/FPMA data, using both the phenomenological DBB+PL and physical TCAF models separately in 
\textit{XSPEC}. Depending on our detailed analysis of GRS 1716-249 during 2016-17 outburst, we may summarize following 
results -

(i) The source showed low frequency QPOs during the rising phase of the outburst. These QPO frequencies were found to 
rise monotonically. QPOs were not originated due to the resonance shock oscillation of the CENBOL.

(ii) The nature of the QPOs are found to be energy dependent. We found that the fundamental QPO is absent above 60 keV, 
while the harmonic is non-detectable above 40 keV in the case of AstroSat/LAXPC data. As the energy increases the rms 
decreases along with the increment of the broadband noise. 

(iii) The $2016-17$ is a `failed' or harder type of outburst which did not go through the usual soft and intermediate spectral 
states. The source was observed only in the hard state (HS) during entire period of the outburst. This non-observation of 
the softer spectral states is similar to what was observed in other short orbital period sources. 

(iv) The probable mass of the black hole was estimated to lie in the range of $4.50-5.93 M_\odot$ 
or $5.02^{+0.91}_{-0.52} M_\odot$, from our spectral analysis with the TCAF model.

\section*{Acknowledgements}

This work has made use of the archival XRT and BAT data provided by UK Swift Science Data Centre at the University of 
Leicester, MAXI GSC data provided by RIKEN, JAXA, and the MAXI team. We have also made use of LAXPC data from Indian 
Space Research Organization's (ISRO's) successful operation of AstroSat mission. We have used FPMA data from NuSTAR 
mission, a project led by Caltech, funded by NASA and managed by NASA/JPL. We have utilised the NuSTARDAS software 
package, jointly developed by the ASDC, Italy and Caltech, USA. 

K.C. acknowledges support from DST/INSPIRE Fellowship (IF170233).
Research of D.D. and S.K.C. is supported in part by the Higher Education Dept. of the Govt. of West Bengal, India.
D.C. and D.D. acknowledge support from DST/SERB sponsored Extra Mural Research project (EMR/2016/003918) fund.
A.J. acknowledges post-doctoral fellowship of Physical Research Laboratory, Ahmedabad, funded by the Department 
of Space, Government of India. 
S.N., D.D. and S.K.C. acknowledge partial support from ISRO sponsored RESPOND project (ISRO/RES/2/418/17-18) fund.
R.B. acknowledges support from CSIR-UGC NET qualified UGC fellowship (June-2018, 527223).

\begin{figure*}
\vskip 0.2cm
  \centering
    \includegraphics[width=10cm]{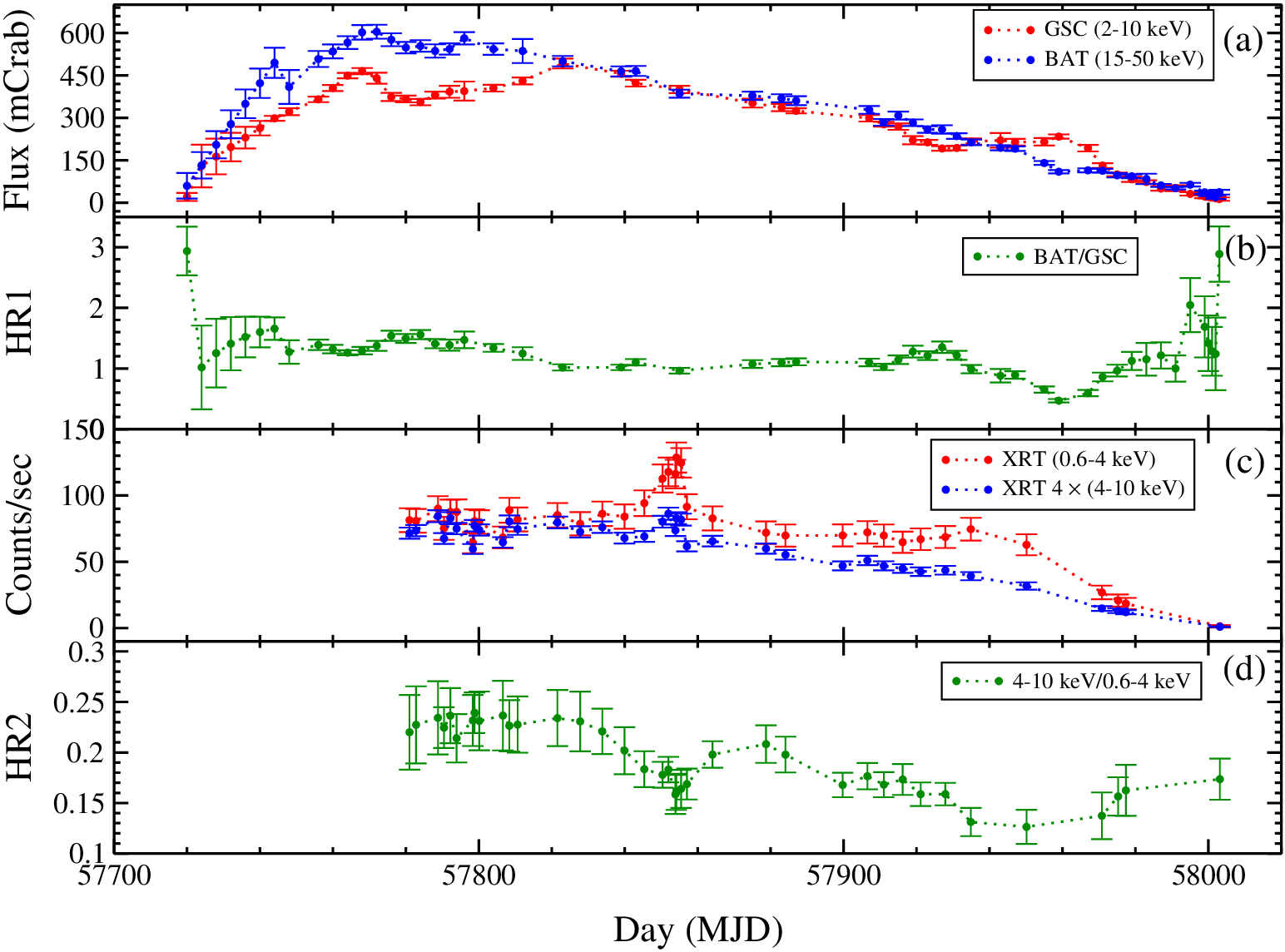}
    \caption{Variation of (a) mCrab converted $15-50$ keV Swift/BAT (online blue) and $2-10$ keV MAXI/GSC (online red) fluxes, (b) hardness ratio (HR) of 
             BAT and GSC fluxes, (c) XRT count rate in $0.6-4$ keV (online red) and $4-10$ keV (online blue) and (d) hardness ratio of $4-10$ keV to $0.6-4$ keV
             XRT count rates throughout the entire $2016-17$ outburst of GRS 1716-249. Note that to scale with the $0.6-4$ keV flux, we have multiplied the 
             $4-10$ keV flux with a factor of 4.}
\end{figure*}

\begin{figure*}
\vskip 0.5cm
  \centering
    \includegraphics[width=07cm,angle=270]{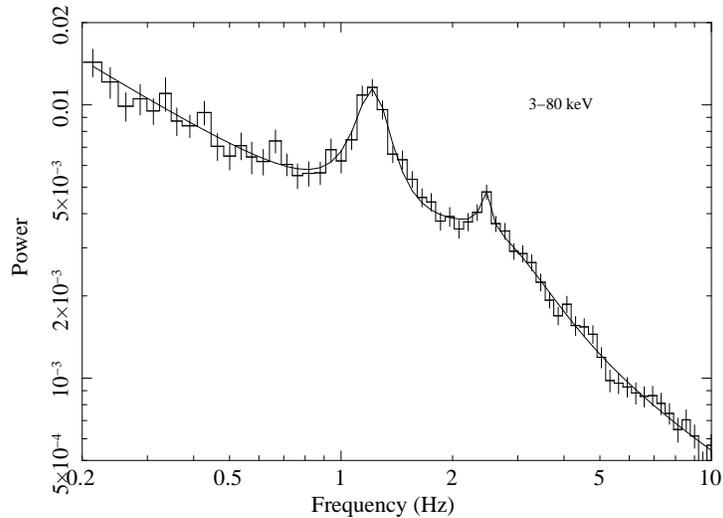}
    \caption{Continuum ($0.2-10$ Hz) fitted power density spectrum (PDS) using 0.01 s time binned $3-80$ keV AstroSat/LAXPC light curve data from the orbit 
            8238 on MJD 57849.69 (2017 April 6). Primary QPO with frequency 1.21 Hz is seen along with its harmonic at 2.42 Hz.}
\end{figure*}

\begin{figure*}
\vskip 0.8cm
  \centering
    \includegraphics[width=10cm]{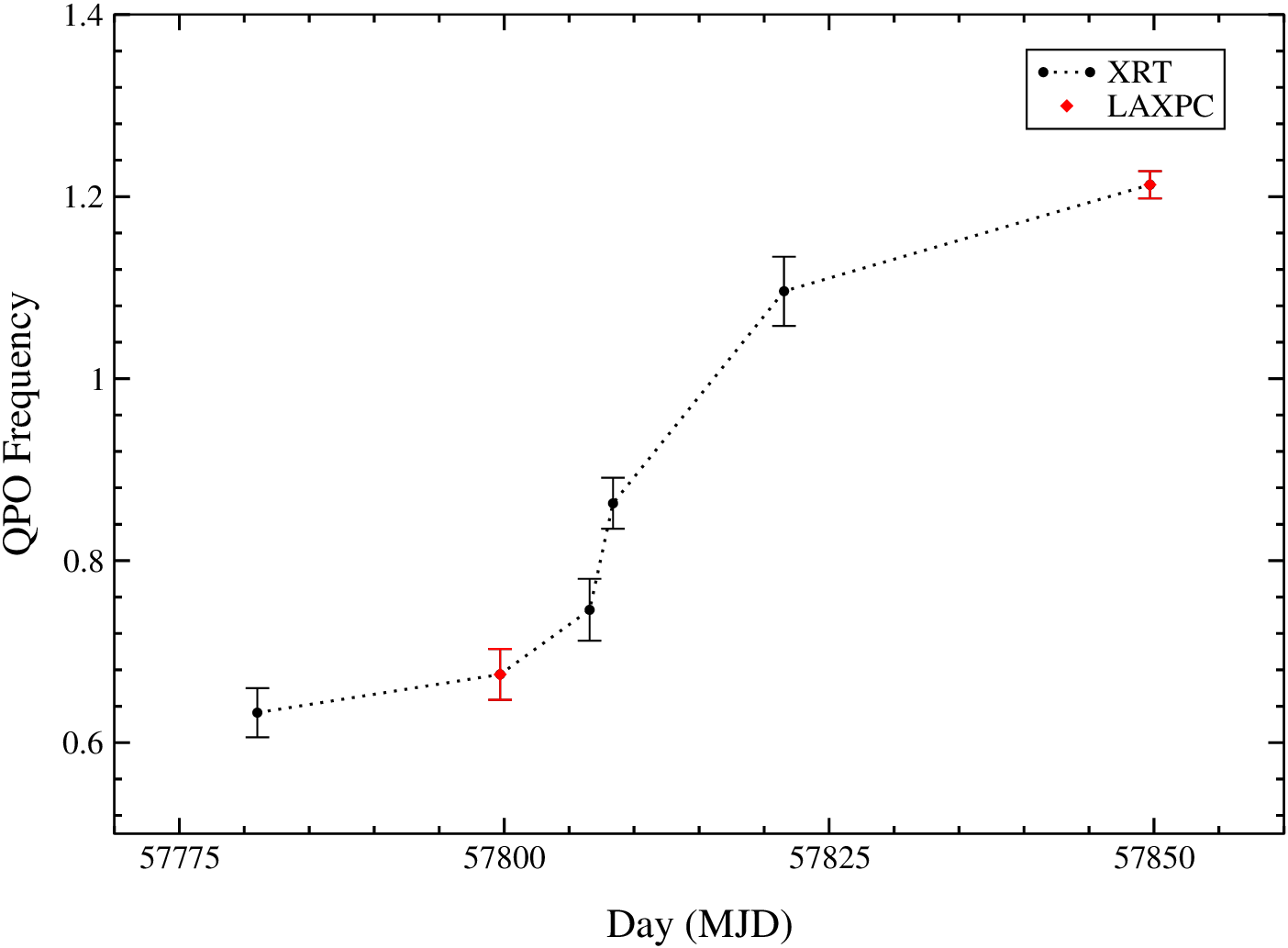}
    \caption{Evolution of QPO frequencies with time (Day in MJD). The black circular points are the ones estimated using Swift/XRT data and red diamond points
             are from AstroSat/LAXPC data. The `y' axis is in units of `Hz'.}
\end{figure*}

\begin{figure*}
\vskip 0.2cm
\centering
\vbox{
\includegraphics[width=4.5truecm,angle=270]{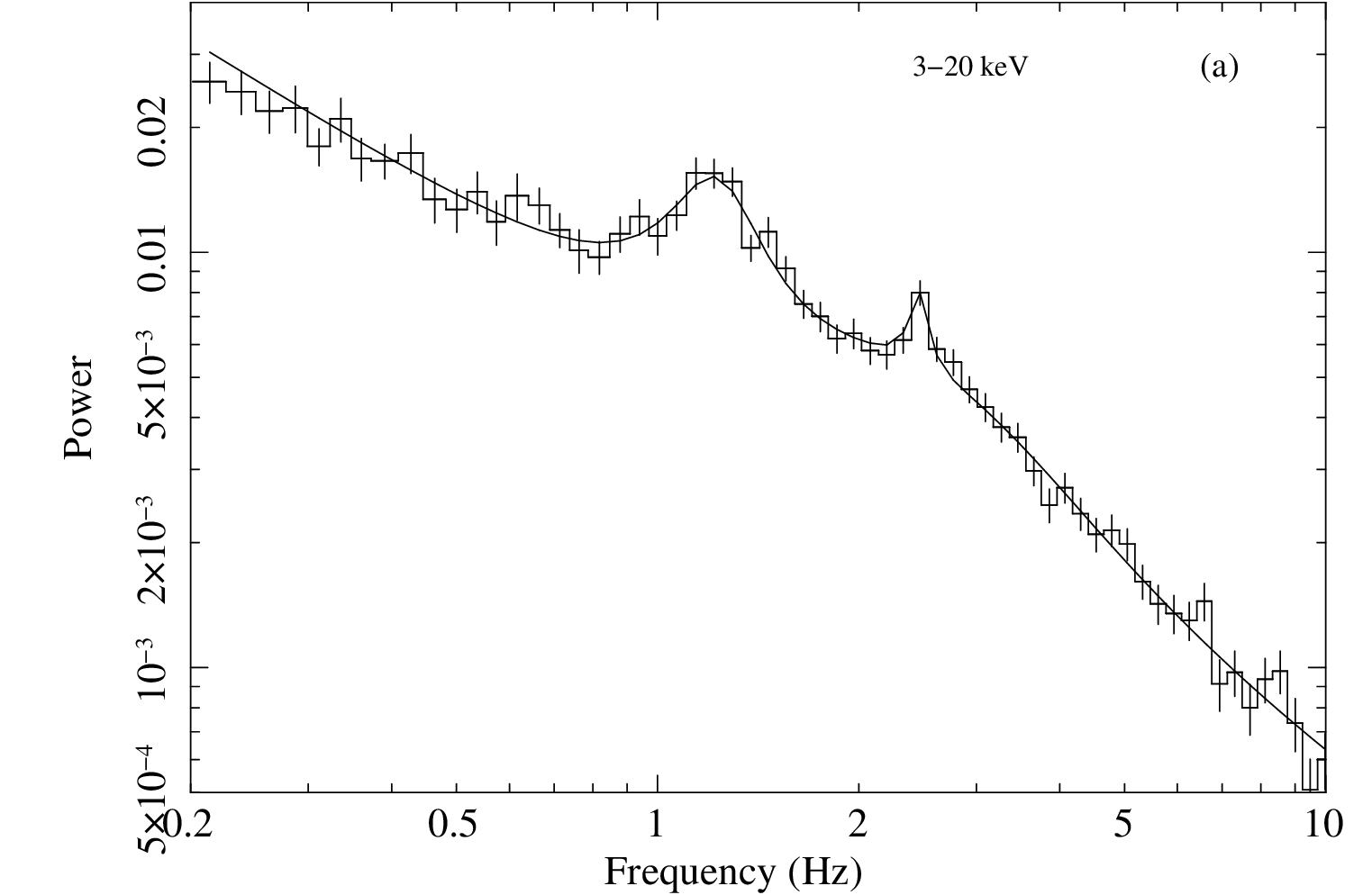}\hskip 0.5cm
\includegraphics[width=4.5truecm,angle=270]{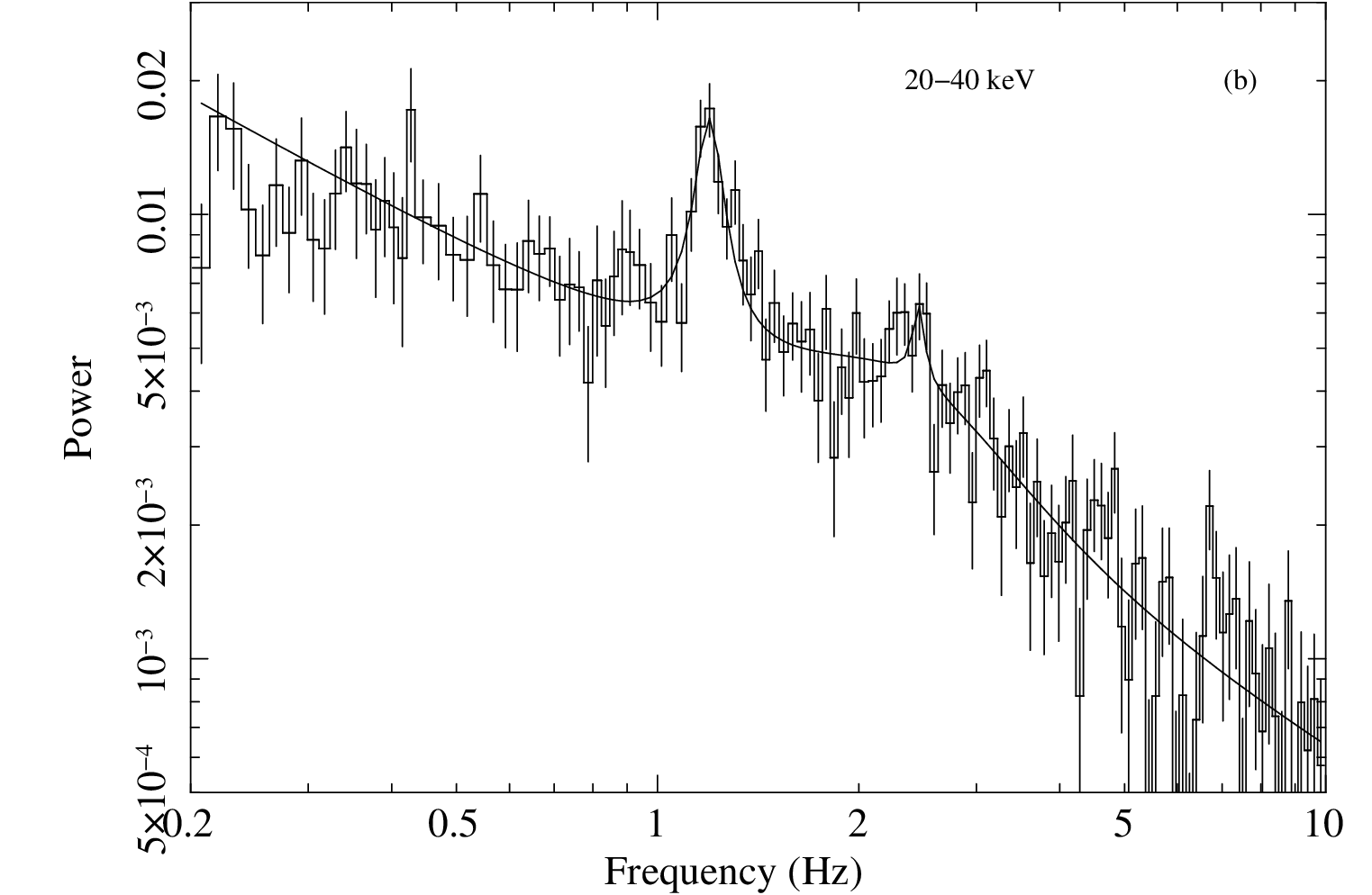}}
\vskip 0.1cm
\hskip 0.5cm
\vbox{
\includegraphics[width=4.5truecm,angle=270]{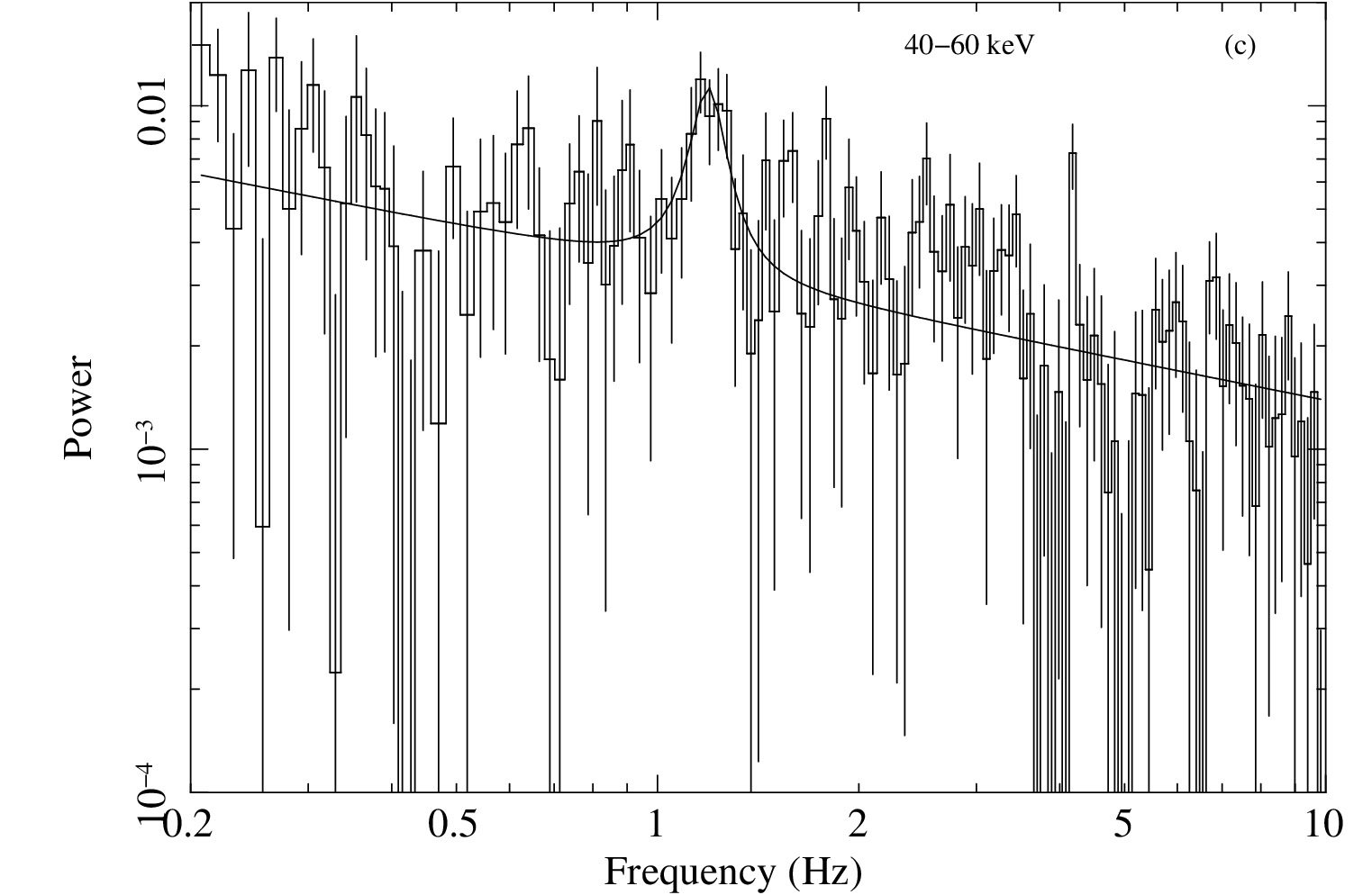}\hskip 0.5cm
\includegraphics[width=4.5truecm,angle=270]{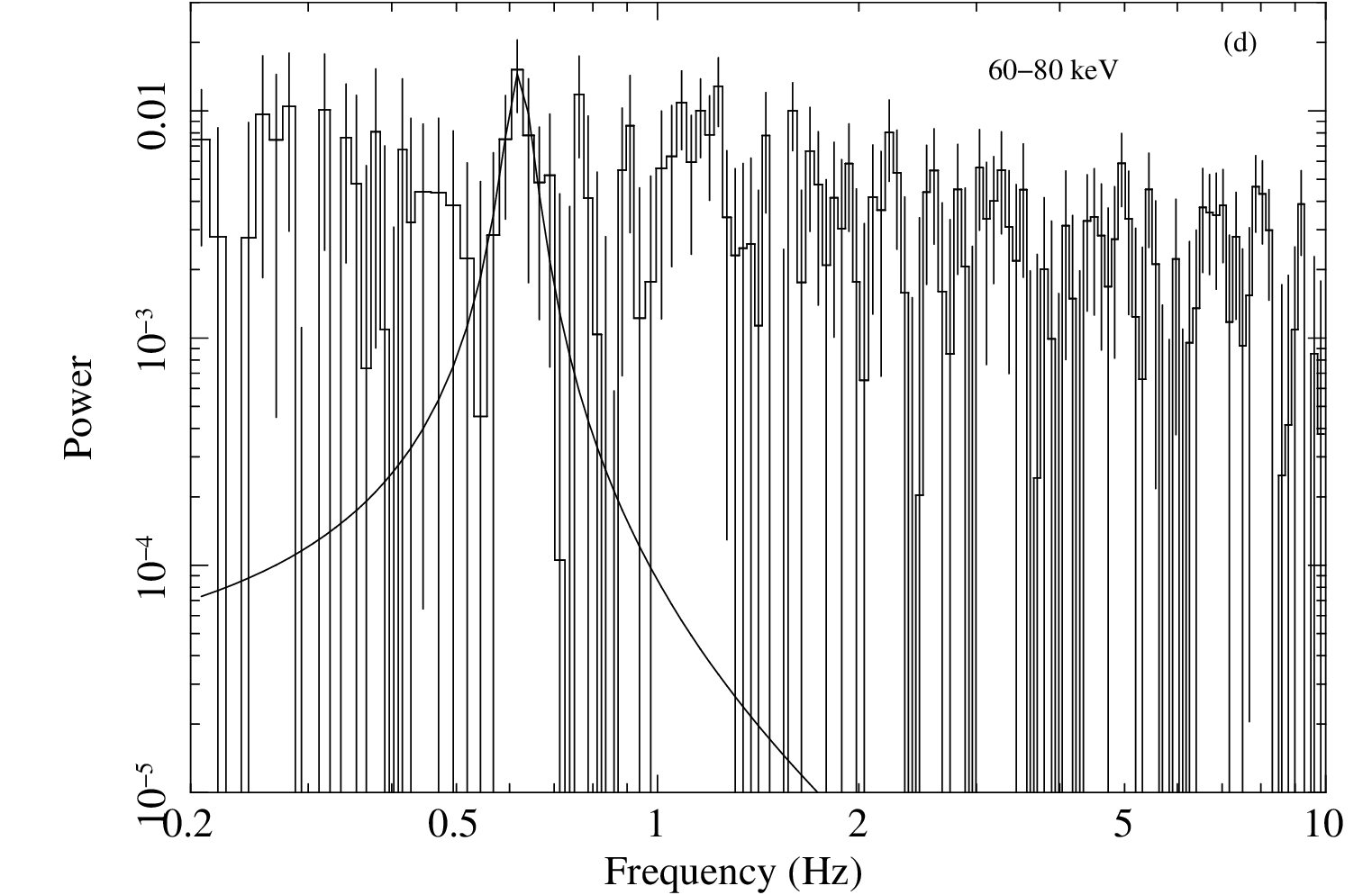} }
\vskip 0.1cm
\caption{AstroSat/LAXPC Continuum ($0.2-10$ Hz) fitted energy dependent power density spectra (PDS) in (a) $3-20$ keV, (b) $20-40$ keV, (c) $40-60$ keV and (d)
         $60-80$ keV energy bands respectively. These PDS are shown for the LAXPC data on 2017 April 06 (MJD 57849.69; obs id = 164T01-9000001140; orbit = 8238).}
\end{figure*}

\begin{figure*}
\vskip 0.2cm
\centering
\vbox{
\includegraphics[width=6.5truecm]{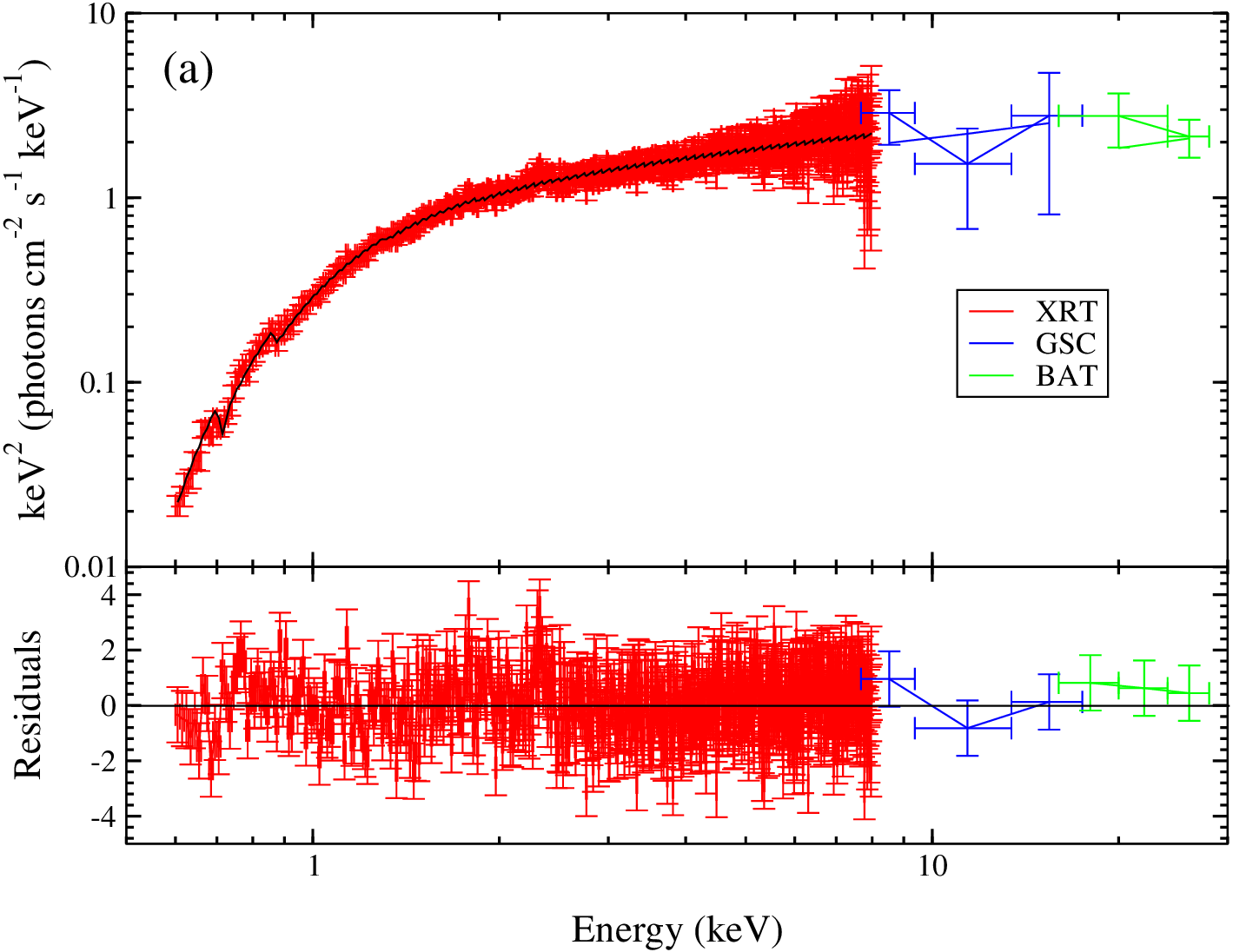}\hskip 0.5cm
\includegraphics[width=6.5truecm]{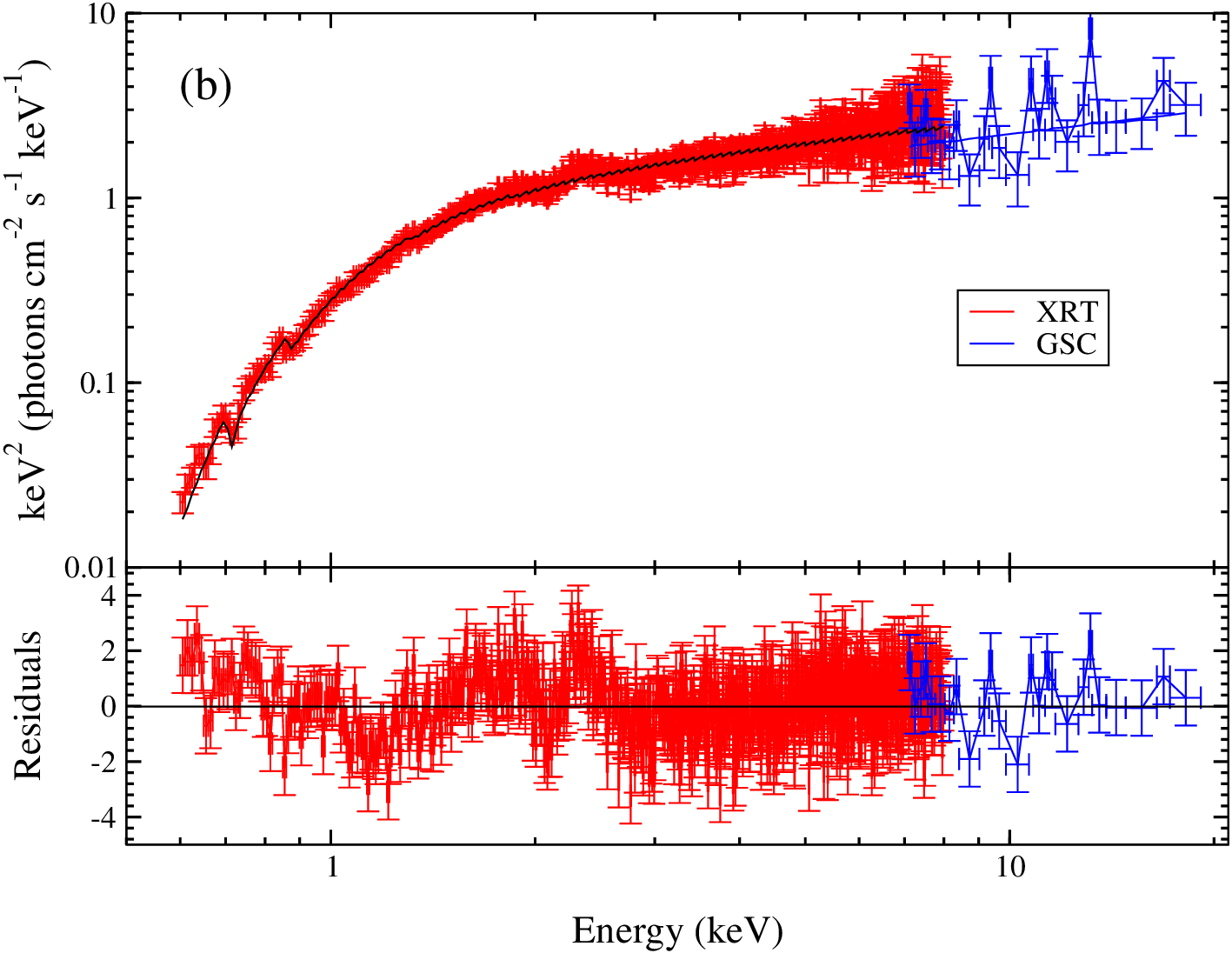}}
\vskip 0.2cm
\hskip 0.5cm
\vbox{
\includegraphics[width=6.5truecm]{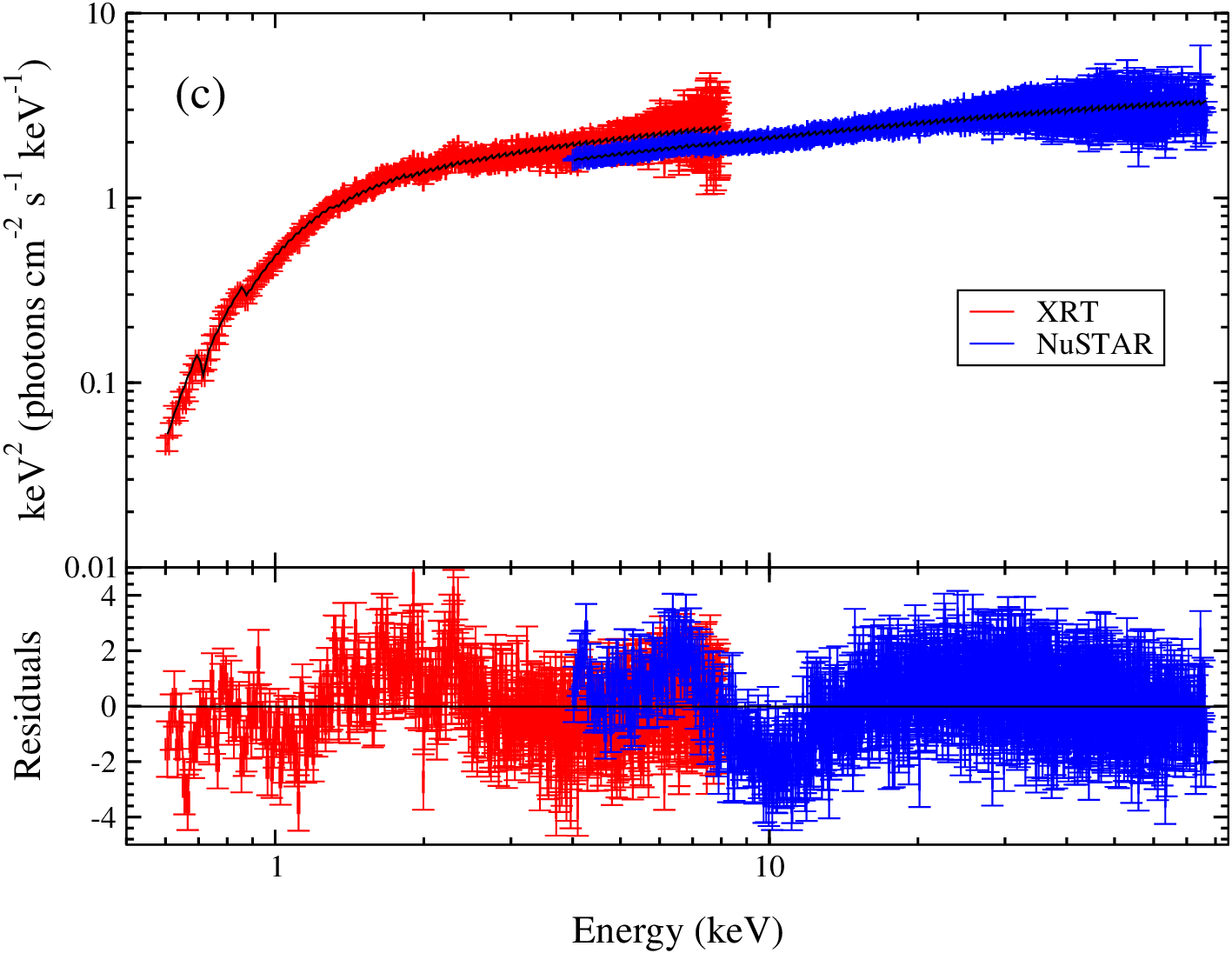}\hskip 0.5cm
\includegraphics[width=6.5truecm]{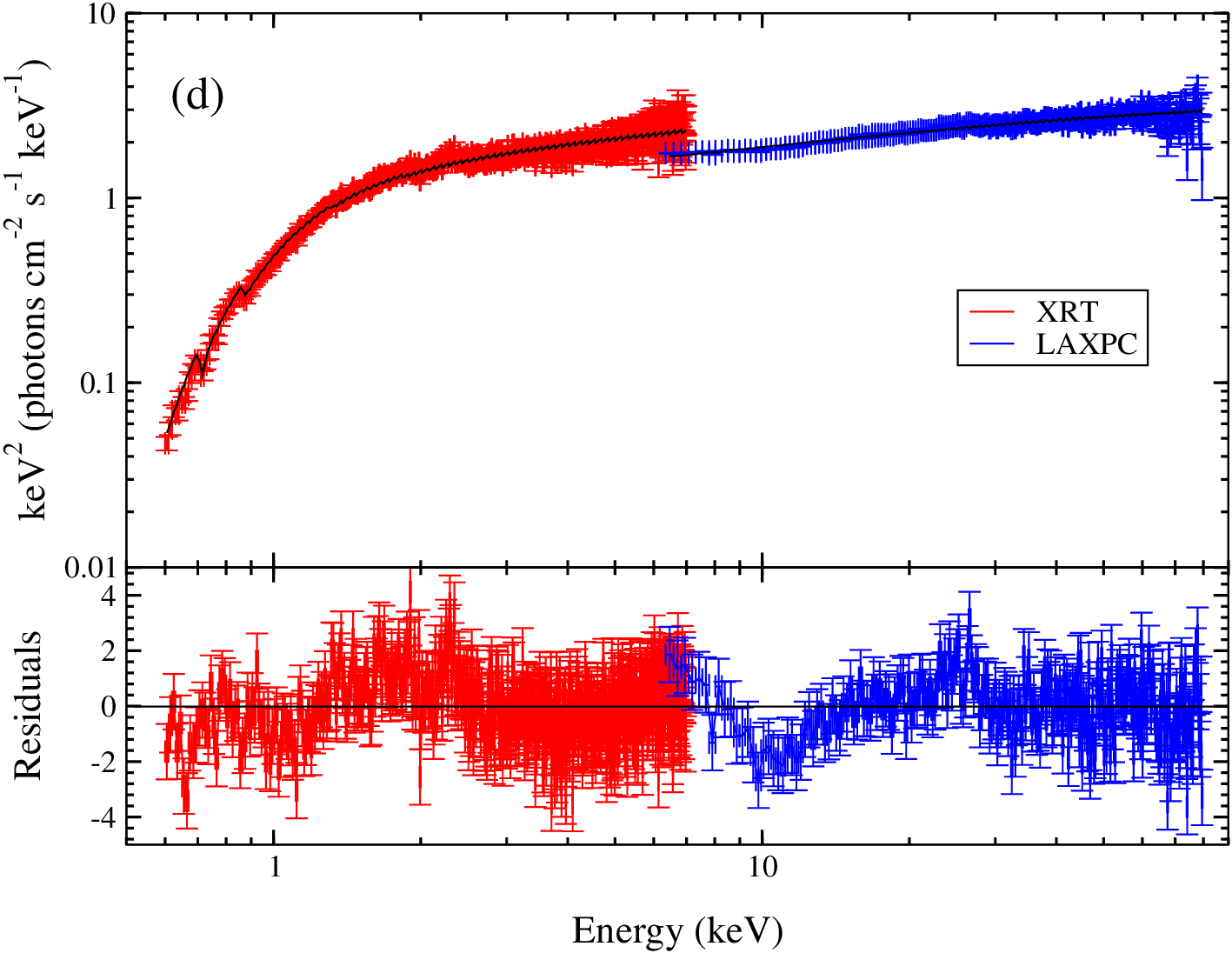} }
\vskip 0.1cm
\caption{TCAF model fitted spectra of differet observations. (a) shows the $0.6-30$ keV XRT+GSC+BAT fitted spectra of the obs. X11 (MJD 57793.56).
         (b) is the $0.6-20$ keV XRT+GSC fitted spectra for obs. X20 (MJD 57808.37). (c) and (d) show broadband spectra using combined XRT+NuSTAR (in 
         $0.6-77$ keV) and XRT+LAXPC (in $0.6-80$ keV) data respectively for XRT obs. X29 (MJD 57850.37). 
\label{spectra}}
\end{figure*}

\begin{figure*}
  \centering
    \includegraphics[width=10cm]{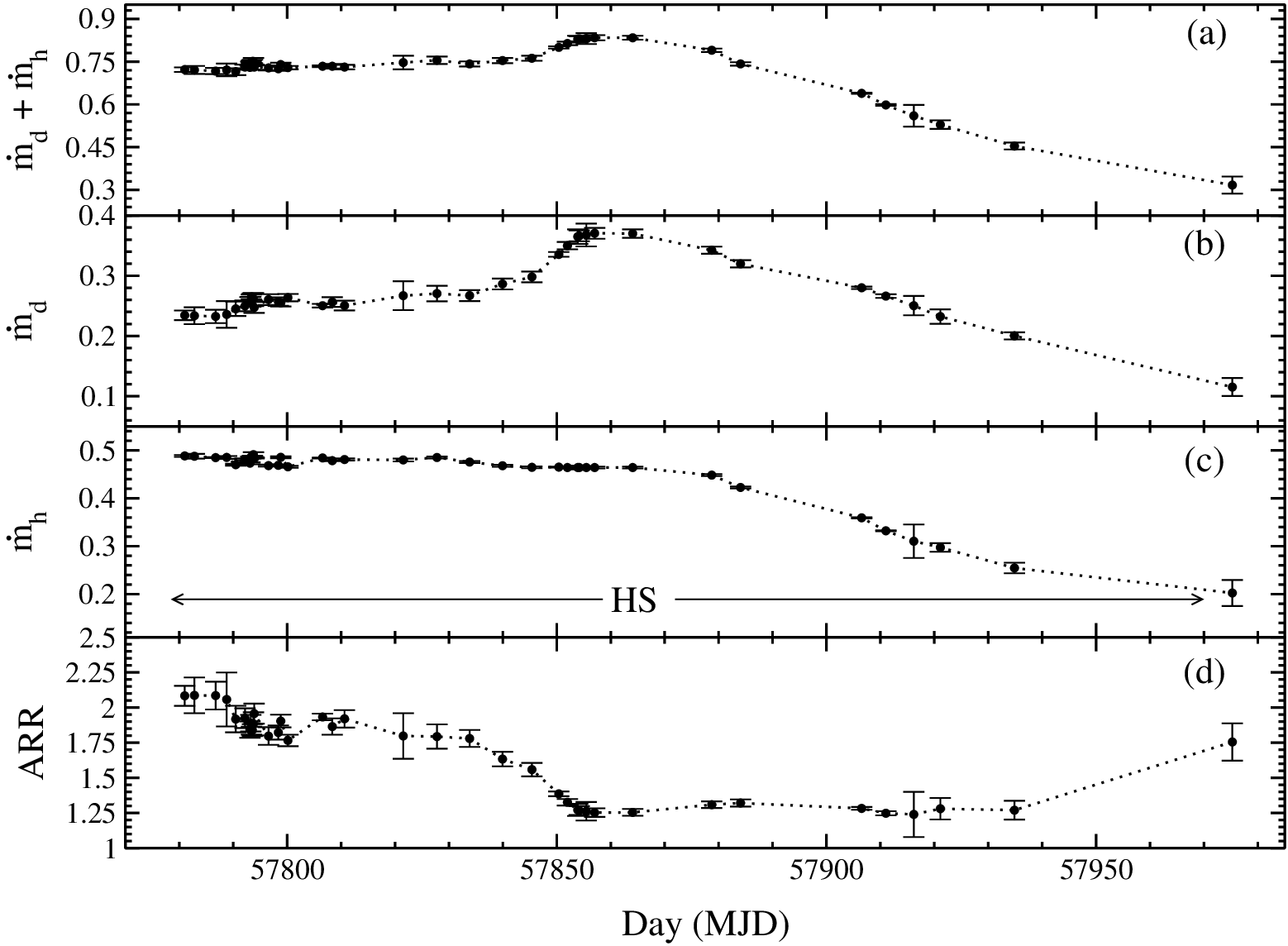}
    \caption{Variation of TCAF model fitted (a) total accretion rate (${\dot m}_d + {\dot m}_h$), (b) disk rate (${\dot m}_d$), (c) halo rate (${\dot m}_h$)
             and (d) accretion rate ratio (ARR = ${\dot m}_h/{\dot m}_d$). The accretion rates are in the unit of Eddington rate ${\dot m}_{Edd}$.}
\end{figure*}

\begin{figure*}
  \centering
    \includegraphics[width=10cm]{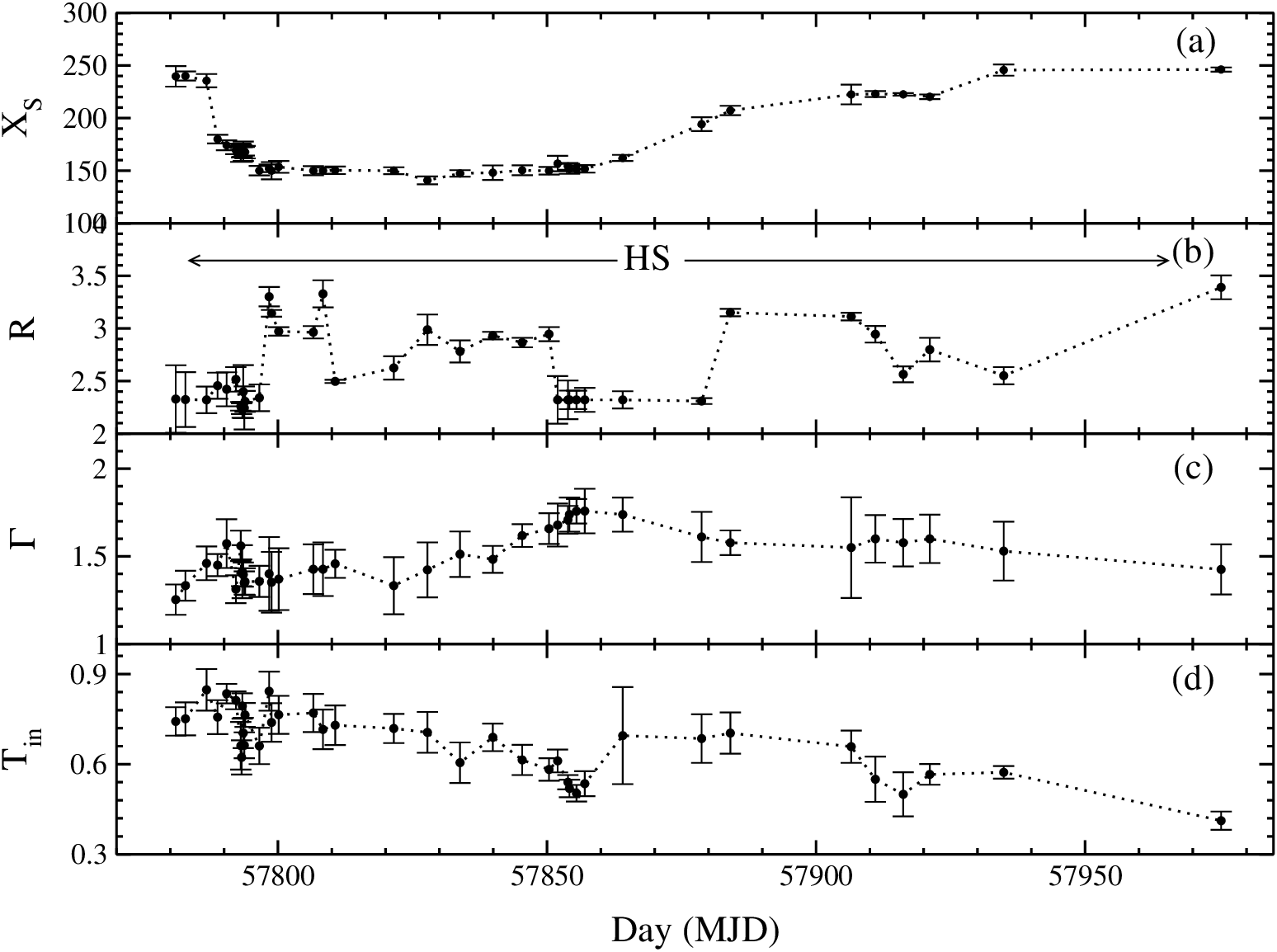}
    \caption{Variations of (a) TCAF model fitted shock location ($X_s$), (b) compression ratio ($R$), (c) DBB+PL fitted photon index ($\Gamma$) and (d) inner
             disk temperature ($T_{in}$) as functions of time (Day in MJD). $X_s$ and $T_{in}$ are in the units of Schwarzschild radius ($r_s$) and keV 
             respectively.}             
\end{figure*}

\begin{figure*}
\vskip 0.5cm
  \centering
    \includegraphics[width=10cm]{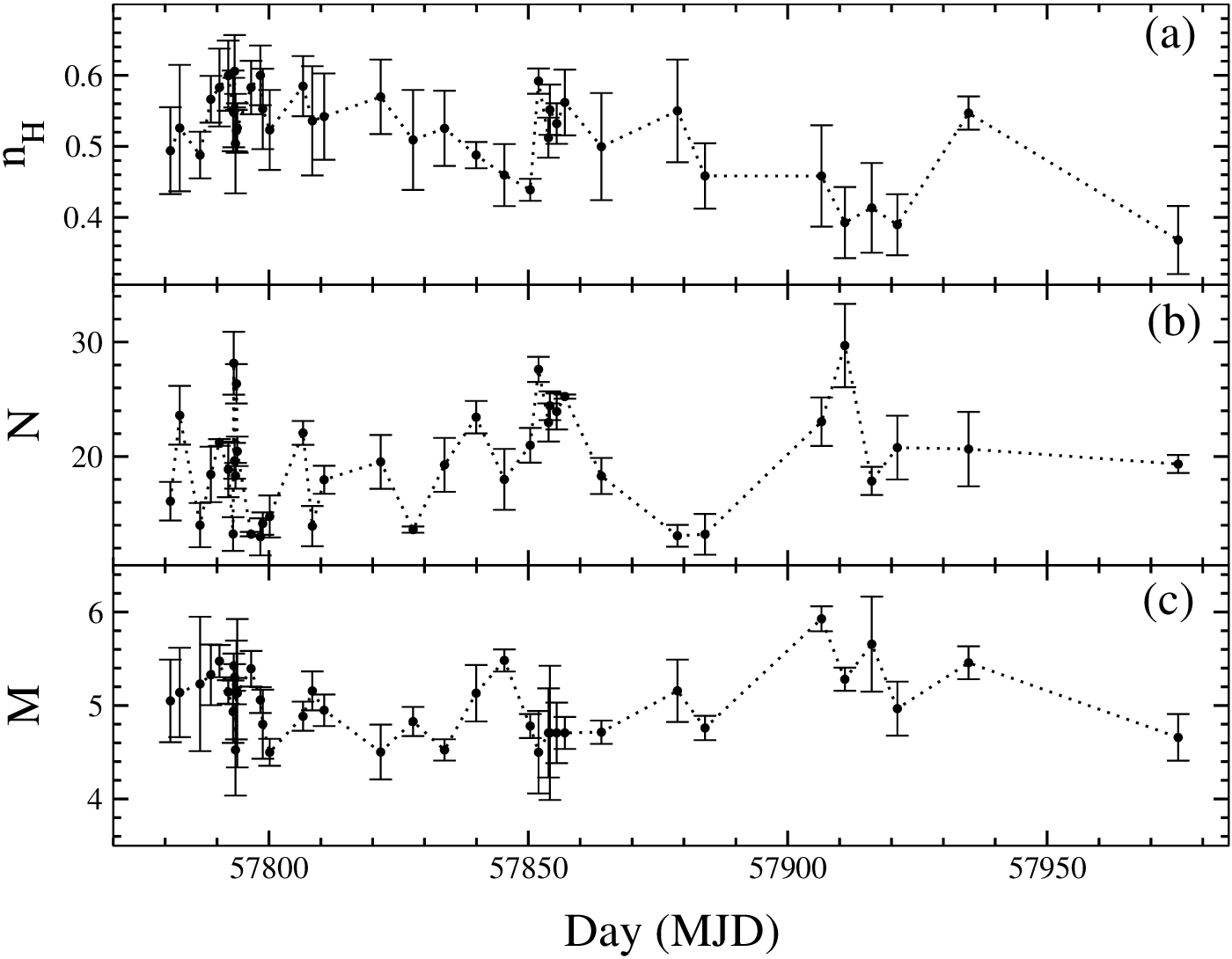}
    \caption{Variations of (a) hydrogen column density ($n_H$), TCAF model fitted (b) normalization ($N$) and 
	mass of the BH ($M_{BH}$ in $M_\odot$) are shown.}             
\end{figure*}

\newcommand{\STAB}[1]{\begin{tabular}{@{}c@{}}#1\end{tabular}}

\begin{table*}
\small
 \addtolength{\tabcolsep}{-3.5pt}
 \centering
 \caption{QPO properties}
 \label{tab:table1}
 \begin{tabular}{lcccccc}
 \hline
UT$^{[1]}$  &    Day$^{[1]}$    & $\nu_{qpo}$$^{[2]}$ &   $\Delta{\nu}$$^{[2]}$ & Q$^{[3]}$   &  rms (\%)$^{[4]}$\\
Date        &       MJD         &      (Hz)           &   (HZ)                  & (${\nu_{qpo}}/{\Delta\nu}$) &    \\
	 (1)      &    (2)         &       (3)         &      (4)            &    (5)                  & (6)      \\
 \hline
 2017-01-28  &  57781.00   &  $0.633 \pm 0.027$   & $0.279 \pm 0.136$   & 2.26  &  14.95  \\
 2017-02-15  &  57799.68   &  $0.675 \pm 0.028$   & $0.371 \pm 0.137$   & 1.81  &  11.60  \\
 2017-02-22  &  57806.57   &  $0.746 \pm 0.034$   & $0.383 \pm 0.178$   & 1.94  &  15.31  \\
 2017-02-24  &  57808.37   &  $0.863 \pm 0.028$   & $0.214 \pm 0.142$   & 4.03  &  12.43  \\
 2017-03-09  &  57821.53   &  $1.096 \pm 0.038$   & $0.432 \pm 0.144$   & 2.53  &  13.53  \\
 2017-04-06* &  57849.69   &  $1.213 \pm 0.015$   & $0.487 \pm 0.081$   & 2.49  &  9.131  \\
\hline
\hline
& \multicolumn{1}{c}{Energy (keV)} &  &  &  & \\ 
\multirow{4}{*}{\STAB{\rotatebox[origin=c]{90}{*LAXPC}}}
         &3-20          &  $1.191 \pm 0.026$   &  $0.601 \pm 0.155$     &  1.98     &     12.03      \\
         &20-40         &  $1.187 \pm 0.014$   &  $0.468 \pm 0.071$     &  2.54     &     11.14      \\
         &40-60         &  $1.191 \pm 0.028$   &  $0.214 \pm 0.087$     &  5.55     &     6.82       \\
         &60-80         &  $0.619 \pm 0.019$   &  $0.058 \pm 0.078$     &  10.66    &     3.66       \\
\hline
 \end{tabular}
\vspace{0.2cm}
 \noindent{
 \leftline{$^{[1]}$ Column 1 and 2 represent the universal time and MJD of the data used. The 2nd and 6th row are 
	the observations, taken using LAXPC data.} 
\leftline{The 2nd data is from the orbit 7498 and last one is from orbit 8238. The 1st, 3rd, 4th and 5th data 
	  are from XRT data on MJD 57781.00 } 
\leftline{(obs id=00034924001), 57806.5 (obs id=00034924019), 57808.3 (obs id=00034924020) and MJD 57821.53 (obs id=00034924022) 
	respectively.}
 \leftline{$^{[2]}$ ${\nu}_{qpo}$ and  $\Delta\nu$ represent observed QPO frequency and its full width at half maximum (FWHM), 
	obtained by fitting PDS with Lorentzian profiles.}
 \leftline{$^{[3]}$ Q (=$\nu_{qpo}$/${\Delta{\nu}}$) is the coherence factor that indicates sharpness of the QPOs.}
 \leftline{$^{[4]}$ Column 7 represents the percentage rms amplitude of the QPOs.}
 \leftline{In the lower panel we show the parameters ($\nu_{qpo}$, $\Delta{\nu}$, Q and rms) for different energy ranges. 
	The results are shown for the LAXPC data on 2017} \leftline{April 06 (orbit=8238), marked with a * in the first panel.}
          }
\end{table*}

\clearpage
\begin{table*}
\small
 \addtolength{\tabcolsep}{-2.5pt}
 \centering
 \caption{Properties of spectral model fitted parameters}
 \label{tab:table2}
\begin{tabular}{ccc|c|ccc|ccccccc}
\hline
  Obs$^{[1]}$ & UT$^{[2]}$ & MJD$^{[2]}$ & $n_H$$^{[3]}$ & T$_{in}$$^{[4]}$ & $\Gamma$$^{[4]}$ & ${\chi}^2$/dof$^{[6]}$ & ${\dot m}_d$$^{[5]}$ & ${\dot m}_h$$^{[5]}$ & $X_s$$^{[5]}$ & $R$$^{[5]}$ & $M_{BH}$$^{[5]}$ & $N$$^{[5]}$ & ${\chi}^2$/dof$^{[6]}$ \\
  (1) & (2) & (3) & (4) & (5) & (6) & (7) & (8) & (9) & (10) & (11) & (12) & (13) & (14) \\
\hline
 X01 & Y-01-28 & 57781.00 & $ 0.49 ^{\pm 0.12} $ & $ 0.74 ^{\pm 0.05} $ & $ 1.25 ^{\pm 0.08} $ & $  776/756 $ & $ 0.234 ^{\pm 0.008} $ & $ 0.488 ^{\pm 0.002} $ & $ 239.7 ^{\pm 9.7} $ & $ 2.32 ^{\pm 0.32} $ & $ 5.05 ^{\pm 0.44} $ & $ 16.1 ^{\pm 1.7} $ & $  804/748 $  \\
 X02 & Y-01-29 & 57782.79 & $ 0.52 ^{\pm 0.12} $ & $ 0.75 ^{\pm 0.05} $ & $ 1.33 ^{\pm 0.08} $ & $  801/748 $ & $ 0.233 ^{\pm 0.014} $ & $ 0.487 ^{\pm 0.005} $ & $ 239.9 ^{\pm 4.3} $ & $ 2.32 ^{\pm 0.26} $ & $ 5.13 ^{\pm 0.47} $ & $ 23.6 ^{\pm 2.6} $ & $  817/752 $  \\
 X04 & Y-02-02 & 57786.73 & $ 0.49 ^{\pm 0.13} $ & $ 0.85 ^{\pm 0.07} $ & $ 1.46 ^{\pm 0.09} $ & $  736/747 $ & $ 0.232 ^{\pm 0.011} $ & $ 0.484 ^{\pm 0.001} $ & $ 235.5 ^{\pm 6.2} $ & $ 2.32 ^{\pm 0.12} $ & $ 5.23 ^{\pm 0.71} $ & $ 14.0 ^{\pm 1.9} $ & $  717/720 $  \\
 X05 & Y-02-04 & 57788.79 & $ 0.57 ^{\pm 0.13} $ & $ 0.76 ^{\pm 0.06} $ & $ 1.45 ^{\pm 0.06} $ & $  935/750 $ & $ 0.235 ^{\pm 0.022} $ & $ 0.485 ^{\pm 0.003} $ & $ 180.0 ^{\pm 4.1} $ & $ 2.45 ^{\pm 0.12} $ & $ 5.32 ^{\pm 0.32} $ & $ 18.4 ^{\pm 2.4} $ & $  925/715 $  \\
 X06 & Y-02-06 & 57790.46 & $ 0.58 ^{\pm 0.25} $ & $ 0.83 ^{\pm 0.32} $ & $ 1.57 ^{\pm 0.13} $ & $  836/742 $ & $ 0.245 ^{\pm 0.012} $ & $ 0.470 ^{\pm 0.002} $ & $ 174.1 ^{\pm 4.7} $ & $ 2.42 ^{\pm 0.16} $ & $ 5.47 ^{\pm 0.17} $ & $ 21.2 ^{\pm 2.2} $ & $  800/709 $  \\
 X07 & Y-02-08 & 57792.18 & $ 0.60 ^{\pm 0.11} $ & $ 0.81 ^{\pm 0.03} $ & $ 1.31 ^{\pm 0.07} $ & $  911/740 $ & $ 0.250 ^{\pm 0.009} $ & $ 0.481 ^{\pm 0.002} $ & $ 169.5 ^{\pm 3.8} $ & $ 2.51 ^{\pm 0.11} $ & $ 5.14 ^{\pm 0.12} $ & $ 18.9 ^{\pm 1.4} $ & $  934/736 $  \\
 X08 & Y-02-09 & 57793.09 & $ 0.55 ^{\pm 0.06} $ & $ 0.66 ^{\pm 0.08} $ & $ 1.55 ^{\pm 0.08} $ & $  783/734 $ & $ 0.256 ^{\pm 0.008} $ & $ 0.473 ^{\pm 0.002} $ & $ 167.1 ^{\pm 8.6} $ & $ 2.25 ^{\pm 0.04} $ & $ 4.93 ^{\pm 0.33} $ & $ 13.2 ^{\pm 1.5} $ & $  790/739 $  \\
 X09 & Y-02-09 & 57793.24 & $ 0.55 ^{\pm 0.15} $ & $ 0.62 ^{\pm 0.06} $ & $ 1.40 ^{\pm 0.07} $ & $  865/739 $ & $ 0.256 ^{\pm 0.008} $ & $ 0.485 ^{\pm 0.002} $ & $ 168.5 ^{\pm 5.9} $ & $ 2.23 ^{\pm 0.05} $ & $ 5.42 ^{\pm 0.13} $ & $ 28.2 ^{\pm 2.7} $ & $  893/738 $  \\
 X10 & Y-02-09 & 57793.37 & $ 0.60 ^{\pm 0.12} $ & $ 0.79 ^{\pm 0.04} $ & $ 1.41 ^{\pm 0.15} $ & $  906/734 $ & $ 0.264 ^{\pm 0.004} $ & $ 0.486 ^{\pm 0.001} $ & $ 167.7 ^{\pm 4.5} $ & $ 2.22 ^{\pm 0.07} $ & $ 5.30 ^{\pm 0.14} $ & $ 19.6 ^{\pm 1.6} $ & $  949/732 $  \\
 X11 & Y-02-09 & 57793.56 & $ 0.50 ^{\pm 0.11} $ & $ 0.70 ^{\pm 0.06} $ & $ 1.39 ^{\pm 0.06} $ & $  837/743 $ & $ 0.264 ^{\pm 0.006} $ & $ 0.487 ^{\pm 0.002} $ & $ 168.4 ^{\pm 9.2} $ & $ 2.39 ^{\pm 0.25} $ & $ 4.52 ^{\pm 0.48} $ & $ 18.3 ^{\pm 1.1} $ & $  854/739 $  \\
 X12 & Y-02-09 & 57793.76 & $ 0.52 ^{\pm 0.13} $ & $ 0.66 ^{\pm 0.04} $ & $ 1.34 ^{\pm 0.06} $ & $  759/734 $ & $ 0.260 ^{\pm 0.011} $ & $ 0.491 ^{\pm 0.005} $ & $ 168.3 ^{\pm 4.0} $ & $ 2.24 ^{\pm 0.20} $ & $ 5.16 ^{\pm 0.52} $ & $ 26.3 ^{\pm 1.7} $ & $  797/734 $  \\
 X13 & Y-02-09 & 57793.89 & $ 0.52 ^{\pm 0.11} $ & $ 0.76 ^{\pm 0.04} $ & $ 1.35 ^{\pm 0.07} $ & $  802/734 $ & $ 0.247 ^{\pm 0.009} $ & $ 0.484 ^{\pm 0.001} $ & $ 167.6 ^{\pm 5.9} $ & $ 2.30 ^{\pm 0.09} $ & $ 5.13 ^{\pm 0.79} $ & $ 20.4 ^{\pm 1.3} $ & $  830/732 $  \\
 X15 & Y-02-12 & 57796.55 & $ 0.58 ^{\pm 0.04} $ & $ 0.66 ^{\pm 0.06} $ & $ 1.35 ^{\pm 0.08} $ & $  779/775 $ & $ 0.260 ^{\pm 0.009} $ & $ 0.468 ^{\pm 0.002} $ & $ 150.0 ^{\pm 4.4} $ & $ 2.34 ^{\pm 0.12} $ & $ 5.39 ^{\pm 0.19} $ & $ 13.2 ^{\pm 1.2} $ & $ 1137/861 $  \\
 X16 & Y-02-14 & 57798.36 & $ 0.60 ^{\pm 0.04} $ & $ 0.84 ^{\pm 0.06} $ & $ 1.40 ^{\pm 0.21} $ & $  855/753 $ & $ 0.257 ^{\pm 0.007} $ & $ 0.468 ^{\pm 0.002} $ & $ 152.2 ^{\pm 3.6} $ & $ 3.30 ^{\pm 0.09} $ & $ 5.05 ^{\pm 0.13} $ & $ 13.0 ^{\pm 1.6} $ & $  941/720 $  \\
 X17 & Y-02-14 & 57798.81 & $ 0.55 ^{\pm 0.06} $ & $ 0.74 ^{\pm 0.06} $ & $ 1.35 ^{\pm 0.17} $ & $  887/738 $ & $ 0.255 ^{\pm 0.006} $ & $ 0.485 ^{\pm 0.002} $ & $ 150.0 ^{\pm 8.2} $ & $ 3.14 ^{\pm 0.03} $ & $ 4.79 ^{\pm 0.36} $ & $ 14.1 ^{\pm 1.1} $ & $  965/777 $  \\
 X18 & Y-02-16 & 57800.14 & $ 0.52 ^{\pm 0.06} $ & $ 0.76 ^{\pm 0.06} $ & $ 1.37 ^{\pm 0.17} $ & $  783/734 $ & $ 0.263 ^{\pm 0.006} $ & $ 0.465 ^{\pm 0.002} $ & $ 153.6 ^{\pm 5.6} $ & $ 2.97 ^{\pm 0.04} $ & $ 4.50 ^{\pm 0.14} $ & $ 14.8 ^{\pm 1.8} $ & $  812/732 $  \\
 X19 & Y-02-22 & 57806.58 & $ 0.58 ^{\pm 0.04} $ & $ 0.77 ^{\pm 0.06} $ & $ 1.42 ^{\pm 0.14} $ & $ 1020/751 $ & $ 0.250 ^{\pm 0.003} $ & $ 0.484 ^{\pm 0.001} $ & $ 150.0 ^{\pm 4.2} $ & $ 2.96 ^{\pm 0.05} $ & $ 4.88 ^{\pm 0.15} $ & $ 22.0 ^{\pm 1.0} $ & $ 1055/749 $  \\
 X20 & Y-02-24 & 57808.37 & $ 0.54 ^{\pm 0.08} $ & $ 0.72 ^{\pm 0.06} $ & $ 1.42 ^{\pm 0.15} $ & $  967/756 $ & $ 0.256 ^{\pm 0.008} $ & $ 0.478 ^{\pm 0.001} $ & $ 150.0 ^{\pm 2.8} $ & $ 3.32 ^{\pm 0.12} $ & $ 5.15 ^{\pm 0.20} $ & $ 13.9 ^{\pm 1.8} $ & $ 1008/756 $  \\
 X21 & Y-02-26 & 57810.64 & $ 0.54 ^{\pm 0.06} $ & $ 0.73 ^{\pm 0.06} $ & $ 1.45 ^{\pm 0.08} $ & $  821/739 $ & $ 0.250 ^{\pm 0.008} $ & $ 0.481 ^{\pm 0.002} $ & $ 150.3 ^{\pm 3.5} $ & $ 2.49 ^{\pm 0.01} $ & $ 4.94 ^{\pm 0.16} $ & $ 18.0 ^{\pm 1.2} $ & $  829/737 $  \\
 X22 & Y-03-09 & 57821.53 & $ 0.57 ^{\pm 0.05} $ & $ 0.72 ^{\pm 0.05} $ & $ 1.33 ^{\pm 0.16} $ & $  912/734 $ & $ 0.267 ^{\pm 0.024} $ & $ 0.480 ^{\pm 0.003} $ & $ 150.0 ^{\pm 3.2} $ & $ 2.62 ^{\pm 0.11} $ & $ 4.50 ^{\pm 0.29} $ & $ 19.5 ^{\pm 2.4} $ & $ 1024/732 $  \\
 X24 & Y-03-15 & 57827.78 & $ 0.51 ^{\pm 0.07} $ & $ 0.71 ^{\pm 0.07} $ & $ 1.42 ^{\pm 0.15} $ & $  965/753 $ & $ 0.270 ^{\pm 0.013} $ & $ 0.485 ^{\pm 0.002} $ & $ 140.8 ^{\pm 3.7} $ & $ 2.98 ^{\pm 0.14} $ & $ 4.82 ^{\pm 0.15} $ & $ 13.6 ^{\pm 1.3} $ & $  972/732 $  \\
 X25 & Y-03-21 & 57833.84 & $ 0.52 ^{\pm 0.05} $ & $ 0.60 ^{\pm 0.07} $ & $ 1.51 ^{\pm 0.12} $ & $  749/734 $ & $ 0.267 ^{\pm 0.009} $ & $ 0.475 ^{\pm 0.002} $ & $ 147.4 ^{\pm 3.0} $ & $ 2.78 ^{\pm 0.10} $ & $ 4.52 ^{\pm 0.11} $ & $ 19.3 ^{\pm 2.3} $ & $  784/732 $  \\
 X26 & Y-03-27 & 57839.94 & $ 0.49 ^{\pm 0.02} $ & $ 0.69 ^{\pm 0.04} $ & $ 1.48 ^{\pm 0.07} $ & $  864/762 $ & $ 0.286 ^{\pm 0.009} $ & $ 0.467 ^{\pm 0.002} $ & $ 148.0 ^{\pm 6.9} $ & $ 2.93 ^{\pm 0.03} $ & $ 5.13 ^{\pm 0.30} $ & $ 23.4 ^{\pm 1.4} $ & $  874/751 $  \\
 X27 & Y-04-02 & 57845.38 & $ 0.46 ^{\pm 0.04} $ & $ 0.61 ^{\pm 0.06} $ & $ 1.61 ^{\pm 0.06} $ & $  811/739 $ & $ 0.298 ^{\pm 0.009} $ & $ 0.464 ^{\pm 0.002} $ & $ 150.3 ^{\pm 4.6} $ & $ 2.86 ^{\pm 0.04} $ & $ 5.48 ^{\pm 0.11} $ & $ 18.0 ^{\pm 2.7} $ & $  861/737 $  \\
 X29 & Y-04-07 & 57850.37 & $ 0.44 ^{\pm 0.02} $ & $ 0.58 ^{\pm 0.04} $ & $ 1.65 ^{\pm 0.08} $ & $  801/740 $ & $ 0.335 ^{\pm 0.004} $ & $ 0.465 ^{\pm 0.001} $ & $ 149.8 ^{\pm 3.5} $ & $ 2.94 ^{\pm 0.06} $ & $ 4.78 ^{\pm 0.12} $ & $ 21.0 ^{\pm 1.5} $ & $  979/740 $  \\
 X30 & Y-04-08 & 57851.96 & $ 0.59 ^{\pm 0.02} $ & $ 0.61 ^{\pm 0.04} $ & $ 1.67 ^{\pm 0.12} $ & $ 1023/742 $ & $ 0.350 ^{\pm 0.006} $ & $ 0.464 ^{\pm 0.002} $ & $ 156.7 ^{\pm 7.3} $ & $ 2.32 ^{\pm 0.22} $ & $ 4.50 ^{\pm 0.44} $ & $ 27.6 ^{\pm 1.1} $ & $ 1114/747 $  \\
 X31 & Y-04-10 & 57853.88 & $ 0.51 ^{\pm 0.03} $ & $ 0.54 ^{\pm 0.02} $ & $ 1.70 ^{\pm 0.07} $ & $  847/742 $ & $ 0.365 ^{\pm 0.012} $ & $ 0.464 ^{\pm 0.004} $ & $ 154.0 ^{\pm 3.2} $ & $ 2.32 ^{\pm 0.18} $ & $ 4.70 ^{\pm 0.47} $ & $ 23.0 ^{\pm 1.7} $ & $ 1136/737 $  \\
 X32 & Y-04-11 & 57854.16 & $ 0.55 ^{\pm 0.04} $ & $ 0.52 ^{\pm 0.03} $ & $ 1.73 ^{\pm 0.09} $ & $  812/738 $ & $ 0.366 ^{\pm 0.009} $ & $ 0.464 ^{\pm 0.001} $ & $ 151.7 ^{\pm 4.7} $ & $ 2.32 ^{\pm 0.08} $ & $ 4.70 ^{\pm 0.71} $ & $ 24.4 ^{\pm 1.2} $ & $ 1032/736 $  \\
 X33 & Y-04-12 & 57855.47 & $ 0.53 ^{\pm 0.03} $ & $ 0.50 ^{\pm 0.03} $ & $ 1.75 ^{\pm 0.06} $ & $  872/745 $ & $ 0.367 ^{\pm 0.019} $ & $ 0.464 ^{\pm 0.003} $ & $ 151.6 ^{\pm 3.0} $ & $ 2.32 ^{\pm 0.08} $ & $ 4.70 ^{\pm 0.32} $ & $ 23.9 ^{\pm 1.6} $ & $ 1126/736 $  \\
 X34 & Y-04-14 & 57857.02 & $ 0.56 ^{\pm 0.05} $ & $ 0.54 ^{\pm 0.04} $ & $ 1.75 ^{\pm 0.12} $ & $  801/739 $ & $ 0.370 ^{\pm 0.009} $ & $ 0.464 ^{\pm 0.002} $ & $ 151.7 ^{\pm 3.5} $ & $ 2.32 ^{\pm 0.11} $ & $ 4.70 ^{\pm 0.17} $ & $ 25.2 ^{\pm 1.2} $ & $  966/737 $  \\
 X35 & Y-04-21 & 57864.07 & $ 0.50 ^{\pm 0.08} $ & $ 0.70 ^{\pm 0.16} $ & $ 1.73 ^{\pm 0.09} $ & $  739/748 $ & $ 0.370 ^{\pm 0.007} $ & $ 0.464 ^{\pm 0.002} $ & $ 162.1 ^{\pm 2.8} $ & $ 2.32 ^{\pm 0.08} $ & $ 4.71 ^{\pm 0.12} $ & $ 18.3 ^{\pm 1.6} $ & $  741/746 $  \\
 X36 & Y-05-05 & 57878.74 & $ 0.55 ^{\pm 0.07} $ & $ 0.68 ^{\pm 0.08} $ & $ 1.61 ^{\pm 0.14} $ & $  829/751 $ & $ 0.342 ^{\pm 0.006} $ & $ 0.448 ^{\pm 0.002} $ & $ 194.2 ^{\pm 6.5} $ & $ 2.30 ^{\pm 0.02} $ & $ 5.15 ^{\pm 0.33} $ & $ 13.1 ^{\pm 1.0} $ & $  828/749 $  \\
 X37 & Y-05-11 & 57884.06 & $ 0.46 ^{\pm 0.16} $ & $ 0.70 ^{\pm 0.07} $ & $ 1.57 ^{\pm 0.07} $ & $  686/740 $ & $ 0.319 ^{\pm 0.006} $ & $ 0.422 ^{\pm 0.002} $ & $ 207.2 ^{\pm 4.4} $ & $ 3.15 ^{\pm 0.03} $ & $ 4.75 ^{\pm 0.13} $ & $ 13.2 ^{\pm 1.8} $ & $  696/737 $  \\
 X40 & Y-06-02 & 57906.52 & $ 0.46 ^{\pm 0.18} $ & $ 0.66 ^{\pm 0.05} $ & $ 1.55 ^{\pm 0.28} $ & $  906/734 $ & $ 0.280 ^{\pm 0.002} $ & $ 0.359 ^{\pm 0.001} $ & $ 222.4 ^{\pm 9.3} $ & $ 3.11 ^{\pm 0.03} $ & $ 5.92 ^{\pm 0.13} $ & $ 23.0 ^{\pm 2.1} $ & $  830/719 $  \\
 X41 & Y-06-07 & 57911.02 & $ 0.39 ^{\pm 0.16} $ & $ 0.55 ^{\pm 0.08} $ & $ 1.60 ^{\pm 0.13} $ & $ 1075/728 $ & $ 0.266 ^{\pm 0.003} $ & $ 0.332 ^{\pm 0.001} $ & $ 222.8 ^{\pm 2.8} $ & $ 2.94 ^{\pm 0.07} $ & $ 5.28 ^{\pm 0.12} $ & $ 29.7 ^{\pm 3.6} $ & $  849/746 $  \\
 X42 & Y-06-12 & 57916.19 & $ 0.41 ^{\pm 0.15} $ & $ 0.50 ^{\pm 0.07} $ & $ 1.57 ^{\pm 0.13} $ & $  979/694 $ & $ 0.250 ^{\pm 0.016} $ & $ 0.310 ^{\pm 0.035} $ & $ 222.5 ^{\pm 1.1} $ & $ 2.56 ^{\pm 0.07} $ & $ 5.65 ^{\pm 0.50} $ & $ 17.9 ^{\pm 1.2} $ & $  882/719 $  \\
 X43 & Y-06-17 & 57921.12 & $ 0.39 ^{\pm 0.12} $ & $ 0.57 ^{\pm 0.03} $ & $ 1.60 ^{\pm 0.13} $ & $ 1068/739 $ & $ 0.232 ^{\pm 0.012} $ & $ 0.297 ^{\pm 0.009} $ & $ 220.2 ^{\pm 2.2} $ & $ 2.79 ^{\pm 0.11} $ & $ 4.96 ^{\pm 0.28} $ & $ 20.8 ^{\pm 2.8} $ & $  835/737 $  \\
 X45 & Y-06-30 & 57934.87 & $ 0.55 ^{\pm 0.16} $ & $ 0.57 ^{\pm 0.02} $ & $ 1.53 ^{\pm 0.16} $ & $  864/739 $ & $ 0.200 ^{\pm 0.006} $ & $ 0.254 ^{\pm 0.011} $ & $ 245.6 ^{\pm 5.3} $ & $ 2.55 ^{\pm 0.08} $ & $ 5.45 ^{\pm 0.17} $ & $ 20.6 ^{\pm 3.2} $ & $  999/736 $  \\
 X56 & Y-08-11 & 57975.27 & $ 0.37 ^{\pm 0.08} $ & $ 0.41 ^{\pm 0.03} $ & $ 1.42 ^{\pm 0.14} $ & $  919/719 $ & $ 0.115 ^{\pm 0.015} $ & $ 0.202 ^{\pm 0.027} $ & $ 246.1 ^{\pm 1.9} $ & $ 3.39 ^{\pm 0.11} $ & $ 4.65 ^{\pm 0.24} $ & $ 19.3 ^{\pm 1.8} $ & $  758/734 $  \\
\hline
\end{tabular}
\vskip 0.2cm
 \noindent{
\leftline{$^{[1]}$ represents observation IDs of the data. Here, `X' marks initial part of the observation IDs: 000349240.}
\leftline{$^{[2]}$ `Y' marks the year of the UT dates, which is 2017. UT dates are in mm/dd format. Column 3 represents the respective MJDs of column 2's dates.}
\leftline{$^{[3]}$ represents the model fitted values of hydrogen column density ($n_H$).}
\leftline{$^{[4]}$ DBB+PL model fitted inner-disk temperature $T_{in}$ (in keV), and PL index ($\Gamma$) are mentioned in Cols. 5-6.}
\leftline{$^{[5]}$ TCAF model fitted parameters: disk rate (${\dot m}_d$ in Eddington rate ${\dot M}_{Edd}$), halo rate (${\dot m}_h$ in ${\dot M}_{Edd}$), shock 
          location, $X_s$ in Schwarzschild radius ($r_s$),} 
\leftline{compression ratio ($R$), mass of the black hole ($M_{BH}$ in solar mass $M_{\odot}$) and normalization ($N$) are mentioned in Cols. 8-13.}
\leftline{$^{[6]}$ DBB+PL and TCAF model fitted ${\chi}^2_{red}$ values are mentioned in column 7 and 14 respectively as ${\chi}^2/dof$, where `dof' represents 
          degrees of freedom.}
\leftline{Note: we present average values of 90\% confidence $\pm$ parameter error values, which are obtained using `err' task in XSPEC.}
\leftline{Note: we use the $\pm$ error as the superscript to save space in the Table}
}
\end{table*}

\begin{table*}
\small
 \addtolength{\tabcolsep}{-4.0pt}
 \centering
 \caption{TCAF model fitted parameters for broadband analysis}
 \label{tab:table3}
\begin{tabular}{ccc|ccc|ccccccc}
\hline
  Obs Id$^{[1]}$ & UT$^{[2]}$ & MJD$^{[2]}$ & Obs Id$^{[3]}$ & UT$^{[4}$ & MJD$^{[4]}$  & ${\dot m}_d$$^{[5]}$ & ${\dot m}_h$$^{[5]}$ & $X_s$$^{[5]}$ & $R$$^{[5]}$ & $M_{BH}$$^{[5]}$ & $N$$^{[5]}$ & ${\chi}^2$/dof$^{[6]}$ \\
  (1) & (2) & (3) & (4) & (5) & (6) & (7) & (8) & (9) & (10) & (11) & (12) & (13) \\
\hline
     &  XRT    &          &             &  NuSTAR    &   & & & & & & & \\
 X29 & Y-04-07 & 57850.36 & 90202055002 & 2017-04-07 & 57850.60 & $0.335 ^{\pm 0.007}$ & $0.464 ^{\pm 0.003}$ &  $141.9 ^{\pm 5.2}$ & $3.53 ^{\pm 0.08}$ &  $4.45 ^{\pm 0.37}$ &  $20.2 ^{\pm 1.0}$ &  2923/2105 \\
 X31 & Y-04-10 & 57853.88 & 90202055004 & 2017-04-10 & 57853.69 & $0.365 ^{\pm 0.009}$ & $0.464 ^{\pm 0.007}$ &  $140.9 ^{\pm 4.1}$ & $3.46 ^{\pm 0.75}$ &  $4.46 ^{\pm 0.73}$ &  $20.3 ^{\pm 1.1}$ &  3053/2035 \\
\hline
     &  XRT    &          &                   &  LAXPC &   & & & & & & & \\
 X16 & Y-02-14 & 57798.36 & 156T01-9000001034 & 2017-02-15 & 57799.68 & $0.256 ^{\pm 0.002}$ & $0.47 ^{\pm 0.005}$ &  $147.9 ^{\pm 4.3}$ &  $3.56 ^{\pm 0.58}$ &  $4.81 ^{\pm 0.47}$ &  $11.5 ^{\pm 0.9}$ &  1193/898 \\
            &            &           &     (orbit=7498)      &            &           &       &        &          &        &        &         &           \\
 X29 & Y-04-07 & 57850.36 & 164T01-9000001140 & 2017-04-06 & 57849.69 & $0.325 ^{\pm 0.004}$ & $0.46 ^{\pm 0.003}$ &  $145.1 ^{\pm 2.6}$ &  $2.70 ^{\pm 0.46}$ &  $4.35 ^{\pm 0.23}$ &  $23.6 ^{\pm 1.1}$ &  1135/891 \\
            &            &           &     (orbit=8238)      &            &           &       &        &          &        &        &         &           \\
\hline
\end{tabular}
\vskip 0.2cm
 \noindent{
\leftline{$^{[1]}$ represents the XRT observation Ids. Here `X' marks the initial part of the obs. Ids: 000349240}
\leftline{$^{[2]}$ represent XRT obs. Ids' respective dates in universal time (yyyy/mm/dd format) and MJDs. `Y' represents the year in each UT.}
\leftline{$^{[3]}$ represents the NuSTAR (row 1 and 2) and LAXPC (row 3 and 4) observation Ids.}
\leftline{$^{[4]}$ represent NuSTAR (row 1 and 2) and LAXPC (row 3 and 4) obs. Ids' respective dates in universal time (yyyy/mm/dd format) and MJDs.}
\leftline{$^{[5]}$ TCAF model fitted parameters: disk rate (${\dot m}_d$ in Eddington rate ${\dot M}_{Edd}$), halo rate (${\dot m}_h$ in ${\dot M}_{Edd}$), 
          shock location, $X_s$ in Schwarzschild radius ($r_s$),} \leftline{compression ratio ($R$), mass of the black hole ($M_{BH}$ in solar mass $M_{\odot}$) 
          and normalization ($N$) are mentioned in Cols. 7-12.}
\leftline{$^{[6]}$ TCAF model fitted ${\chi}^2_{red}$ values are mentioned in column 13 as ${\chi}^2/dof$, where `dof' represents degrees of freedom.}
\leftline{Note: we present average values of 90\% confidence $\pm$ parameter error values, which are obtained using `err' task in XSPEC.}
\leftline{Note: we use the $\pm$ error as the superscript to save space in the Table}
}
\end{table*}

\end{document}